\documentclass[acmtosem,manuscript]{acmart}
\AtBeginDocument{%
  }
\setcopyright{none}
\renewcommand{\footnotetextcopyrightpermission}[1]{}

\usepackage{xcolor}
\usepackage{enumitem}
\usepackage{amsmath}
\usepackage{adjustbox}
\usepackage{xspace}
\usepackage{subcaption}
\usepackage{listings}
\usepackage{tcolorbox}
\usepackage{xcolor}

\newcommand{\ie}{\textit{i.e.,}\xspace}
\newcommand{\eg}{\textit{e.g.,}\xspace}
\newcommand{\darkred}{\color[RGB]{139,0,0}}
\newcommand{\darkgreen}{\color[RGB]{0,100,0}}
\definecolor{darkgreen}{rgb}{0.0, 0.5, 0.0}
\definecolor{boxcolor}{RGB}{242,242,242}
\definecolor{bordercolor}{RGB}{0,0,0}

\newcommand{\toolname}{\textsc{EvaLooop}\xspace}

\usepackage{fontawesome5}

\usepackage{booktabs}
\usepackage{longtable} 
\usepackage{caption}   

\usepackage{tabularx}
\usepackage{adjustbox}
\usepackage{multicol}

\usepackage{tikz}
\newcommand*\circled[1]{\tikz[baseline=(char.base)]{
            \node[shape=circle,draw,inner sep=2pt] (char) {#1};}}

\usepackage{soulutf8}
\usepackage{geometry}
\usepackage{setspace}

\tcbuselibrary{theorems,skins,breakable,listings}

\definecolor{myyellow}{rgb}{1,1,0.8}
\sethlcolor{myyellow}
\definecolor{dkgreen}{rgb}{0,0.6,0}
\definecolor{mauve}{rgb}{0.58,0,0.82}
\definecolor{framegray}{gray}{0.8}

\newtcbtheorem[auto counter, number within=section]{myCodeFrame}{Listing}{
  enhanced, verbatim,
  colback=white, colframe=black, boxrule=0.5pt, arc=0.3mm,
  left=1mm, right=1mm, top=0.5mm, bottom=0.5mm,
  fonttitle=\bfseries,
  fontupper=\small\ttfamily\linespread{0.85}\selectfont,
  before upper=\vspace{-2pt},   
  after  upper=\vspace{-2pt},   
  before skip=6pt,              
  after  skip=6pt
}{lst}

\begin{document}

\title{\toolname: A Self-Consistency-centered Framework for Assessing Large Language Model Robustness in Programming}

\author{Sen Fang}
\email{sfang9@ncsu.edu}
\affiliation{%
  \institution{North Carolina State University}
  \city{Raleigh}
  \state{North Carolina}
  \country{USA}
}
\author{Weiyuan Ding}
\email{wding8@ncsu.edu}
\affiliation{%
  \institution{North Carolina State University}
  \city{Raleigh}
  \state{North Carolina}
  \country{USA}
}
\author{Mengshi Zhang}
\email{mengshiz@tensorblock.co}
\affiliation{%
  \institution{TensorBlock}
  \city{Newark}
  \state{Delaware}
  \country{USA}
}
\author{Zihao Chen}
\email{zihaoc@tensorblock.co}
\affiliation{%
  \institution{TensorBlock}
  \city{Newark}
  \state{Delaware}
  \country{USA}
}
\author{Bowen Xu}
\email{bxu22@ncsu.edu}
\affiliation{%
  \institution{North Carolina State University}
  \city{Raleigh}
  \state{North Carolina}
  \country{USA}
}


\begin{abstract}
    \textbf{Motivation:} Evaluating the programming robustness of large language models (LLMs) is paramount for ensuring their reliability in AI-based software development. 
    However, adversarial attacks exhibit fundamental limitations that compromise fair robustness assessment: 
    they demonstrate contradictory evaluation outcomes where different attack strategies tend to favor different models, 
    and more critically, 
    they operate solely through external perturbations, failing to capture the intrinsic stability essential for autonomous coding agents where subsequent inputs are endogenously generated by the model itself.\\
    \textbf{Solution:} We introduce \toolname, a novel assessment framework that evaluates robustness from a self-consistency perspective, leveraging the natural duality inherent in software engineering tasks (\eg code generation and code summarization). 
    \toolname establishes a self-contained feedback loop where an LLM iteratively transforms between code and natural language until functional failure occurs, with robustness quantified by a novel Average Sustainable Loops (ASL) metric—the mean number of iterations maintaining functional correctness across benchmark tasks. This cyclical strategy intrinsically evaluates robustness without relying on external attack configurations, providing a unified metric that reveals how effectively LLMs preserve semantic integrity through sustained self-referential transformations.\\
    \textbf{Results:} We evaluate 96 popular LLMs, ranging from 0.5B to 685B parameters, on \toolname equipped with the MBPP Plus benchmark, and found that \toolname typically induces a 2.65\%–47.62\% absolute drop in pass@1 accuracy within ten loops. 
    Intriguingly, robustness does not always align with initial performance (\ie one-time query); 
    for instance, Qwen3-235B-A22B-Instruct-2507, despite inferior initial code generation compared to OpenAI's o-series models and DeepSeek-V3, demonstrated the superior robustness (ASL score).
    \textbf{Impact:}
    \toolname reveals distinct degradation patterns across different models, providing developers with a practical framework for evaluating LLM functional coherence through iterative transformations and offering complementary insights that support more informed model selection decisions.
    \textbf{A living leaderboard: \url{https://evalooop.github.io/}.}
\end{abstract}


\maketitle


\section{Introduction}

Robustness is a critical aspect for measuring the performance of Large Language Models (LLMs) in code generation~\cite{chen2021evaluating, roziere2023code, nijkamp2023codegen, lozhkov2024starcoder, achiam2023gpt, liu2024deepseek}. 
The reasons are threefold. 
First, real-world consequences of non-robust code generation can be severe: 
Research demonstrates that 11.56\% of websites using LLM-generated PHP code could be compromised, with 26\% having at least one exploitable vulnerability~\cite{dora2025hidden}, 
while Meta's CyberSecEval studies reveal that approximately one in three AI-generated code pieces contain vulnerabilities~\cite{bhatt2023purple, bhatt2024cyberseceval, wan2024cyberseceval}. Production incidents have included SQL injection vulnerabilities from string concatenation patterns, exposure of hard-coded API keys, and OS command injection flaws introduced through AI-suggested code~\cite{mohsin2024can, yan2025guiding}. 
Second, modern software development increasingly relies on LLM-powered tools: Studies show that LLM-generated code has become integral to contemporary development workflows, with developers increasingly accepting AI-assisted code suggestions despite inherent security risks~\cite{mastropaolo2023robustness, rasnayaka2024empirical, jimenez2023swe}. 
Third, the shift toward agent-based software engineering means that LLMs must maintain consistency across multiple interconnected tasks, where cascading inconsistencies can propagate through entire systems. Recent studies note instances of message-code inconsistency, such as ``Phantom Changes'' where descriptions may diverge from the actual code implementation~\cite{gong2026analyzing}. Together with challenges in code-comment alignment~\cite{10416264}, such semantic drifts present a potential hurdle for the reliability of autonomous agents.

Evaluating the robustness of LLMs in programming often involves adversarial attacks~\cite{anand2021adversarial,wang2022recode, jha2023codeattack, nguyen2023adversarial, chen2023evaluating, zhang2024attacks, improta2025enhancing, wei2023towards, zaremba2025trading, bavcic2024towards}.
However, these approaches exhibit fundamental limitations since they primarily focus on robustness against \textit{external} perturbations, failing to capture the \textit{intrinsic} semantic instability that leads to the aforementioned inconsistencies (e.g., phantom changes) in autonomous generation. 
Even more critically, we find that existing works on evaluating LLM robustness are often biased due to their reliance on different adversarial attack strategies. 
For example, when we apply three adversarial attacks~\cite{sclar2024quantifying}, \ie Case Transformation, Structural Reformatting, and Redundant Elaboration, to evaluate the robustness of deepseek-coder-7b-instruct-v1.5 and Qwen2.5-Coder-
32B-Instruct, 
we observe completely contradictory robustness patterns: 
deepseek-coder-7b-instruct-v1.5 demonstrates severe vulnerability to structural attacks (performance drops from 64.6\% to 1.1\%) while maintaining robustness under verbose prompt attacks, 
whereas Qwen2.5-Coder-32B-Instruct shows the opposite behavior—resilience to structural changes but significant degradation under verbose prompts (76.2\% to 54.5\%). 
As a result, both model developers and consumers could be easily misled depending on which attack methodology is employed for evaluation.

From the software engineering workflow perspective, agent-based frameworks represent the trending solution for complex programming tasks~\cite{zhang2024autocoderover, chen2025prometheus, xia2025demystifying}. 
In these frameworks, one task's output frequently serves as another task's input, with each task handled by an LLM agent operating in sequence or parallel configurations. 
Notably, these agent systems typically employ post-trained LLMs that have undergone reinforcement learning~\cite{rafailov2023direct, shao2024deepseekmath} or instruction-based fine-tuning~\cite{ouyang2022training, dubey2024llama} to enhance their reliability and instruction-following capabilities. However, our empirical analysis reveals that such post-trained models demonstrate significant resilience to adversarial attacks, rendering traditional adversarial evaluation approaches largely ineffective for assessing their robustness. 
This creates a critical evaluation limitation: 
while adversarial attacks fail to meaningfully stress-test these resilient models, agent deployments require assessing a fundamentally different dimension of robustness—the model's ability to maintain functional coherence through sustained self-referential transformations where subsequent inputs are endogenously generated by the model itself rather than provided by external adversaries. 
\textbf{These gaps underscore the urgent need for a unified approach for LLMs robustness evaluation in programming that does not depend on external attack configurations.}

\textbf{Our solution.} Inspired by cognitive resilience principles in human learning, we introduce \toolname, a novel framework for evaluating LLM robustness in programming that addresses these fundamental limitations. 
Drawing from the Levels of Processing theory~\cite{craik2002levels}, deep cognitive engagement—such as achieving bilingual fluency or mastering complex dual relationships—builds robust knowledge structures that are inherently resistant to minor perturbations. 
Modern LLMs, having been trained on vast code corpus, demonstrably acquiring sophisticated coding capabilities and constructed resilient knowledge representations that should theoretically withstand iterative transformations.

Specifically, \toolname practices this principle by leveraging the natural duality~\cite{wei2019code} between complementary software engineering tasks (e.g., code generation~\cite{li2022competition} and code summarization~\cite{ahmad2020transformer}) to create a self-contained feedback loop. 
Take the popular dual tasks, code generation and code summarization, as an example, starting from an initial natural language specification, the framework instructs an LLM to generate corresponding code, then tasks the same model with summarizing this generated code back into a new natural language specification. 
This cyclical process continues iteratively until the generated code fails to pass functional correctness tests, with robustness quantified by our proposed Average Sustainable Loops (ASL) metric—the mean number of successful iterations before functional failure occurs.
\toolname essentially constitutes a minimal agent-based system where each loop represents an autonomous agent interaction. Higher ASL scores indicate superior self-consistency, suggesting that the LLM is better suited as a foundation model for complex agent-based software systems. More crucially, \toolname eliminates all external perturbations—beyond the initial input, all subsequent inputs are endogenously generated by the model itself—thereby providing a fair, unbiased comparison framework that captures the intrinsic stability essential for real-world deployment scenarios.

\begin{figure*}[th]
\begin{center}
\begin{minipage}[t]{\textwidth}
\begin{myCodeFrame}{The initial solution generated by o3-mini. \faThumbsUp}{ini_solution}
\scriptsize
Initial prompt:
Write a python function to determine whether a given list of numbers forms a full sequence of
consecutive integers without any duplicates.
\begin{lstlisting}[language=Python, escapeinside={(*}{*)}, basicstyle=\scriptsize\ttfamily]
assert check_Consecutive([1,2,3,4,5]) == True
\end{lstlisting}

Solution:
\begin{lstlisting}[language=Python, escapeinside={(*}{*)}, basicstyle=\scriptsize\ttfamily]
def check_Consecutive(lst):
    if not lst:
        return False
    (*\hl{s = set(lst)}*)
    (*\hl{return max(lst) - min(lst) + 1 == len(lst) and s == set(range(min(lst), max(lst) + 1))}*)
\end{lstlisting}
\end{myCodeFrame}
\end{minipage}


\begin{minipage}[t]{\textwidth} 
\begin{myCodeFrame}{The solution generated by o3-mini in 5-th loop. \faThumbsDown}{failed_solution}
\scriptsize
Derived prompt:
Write a python function to determine if a list of numbers contains an unbroken sequence of consecutive
integers from its minimum to its maximum value.
\begin{lstlisting}[language=Python, escapeinside={(*}{*)}, basicstyle=\scriptsize\ttfamily]
assert check_Consecutive([1,2,3,4,5]) == True
\end{lstlisting}

Solution:
\begin{lstlisting}[language=Python, escapeinside={(*}{*)}, basicstyle=\scriptsize\ttfamily]
def check_Consecutive(lst):
    if not lst:
        return False
    (*\hl{return set(lst) == set(range(min(lst), max(lst) + 1))}*)

assert check_Consecutive([1,2,3,4,5]) == True
\end{lstlisting}
\end{myCodeFrame}

\end{minipage}
\vspace{-6mm}
\end{center}
\caption{The generated solutions for MBPP$^+$ No. 472 task by o3-mini.}
\label{fig:motivation-exp}
\vspace{-4mm}
\end{figure*}

%

Figure~\ref{fig:motivation-exp} demonstrates \toolname's effectiveness through OpenAI o3-mini's performance on MBPP Plus task No.472. Initially, the model generates a correct solution that properly checks both consecutive sequence formation and duplicate absence (Listing~\ref{lst:ini_solution}). However, by the 5th loop, despite receiving a semantically equivalent prompt derived from its own previous output, o3-mini produces a fundamentally flawed implementation (Listing~\ref{lst:failed_solution}) that fails to verify duplicate absence—erroneously returning True for inputs containing duplicates.
This degradation exemplifies the critical self-consistency failures that adversarial attacks cannot capture. While external perturbations might leave this model seemingly robust, \toolname reveals how the same LLM loses functional coherence through endogenous transformations—precisely the type of cascading failures that compromise agent-based systems in practice. Such intrinsic instability, measurable only through sustained self-referential evaluation, directly impacts the reliability of autonomous coding agents where one model's output becomes another's input across multiple reasoning steps.

\textbf{Contributions.} Our main contributions are:
\begin{itemize}[wide, labelwidth=0em, labelindent=0pt]
    \item \textbf{Problem Identification with Empirical Evidence:} We systematically identify and empirically validate critical limitations of adversarial attack-based robustness evaluation methods through comprehensive experiments across 13 state-of-the-art LLMs. 
    Our findings reveal that these approaches exhibit negligible effectiveness against modern post-trained models (e.g., OpenAI's o-series) and demonstrate inherent evaluation bias where different attack strategies tend to favor different models, leading to contradictory robustness conclusions that compromise fair comparison and practical applicability.

    \item \textbf{Novel Evaluation Framework (\toolname):} We propose \toolname, a self-consistency framework that leverages the natural duality between code generation and code summarization within a self-contained feedback loop to provide unified and unbiased evaluation. 
    Our proposed Average Sustainable Loops (ASL) metric quantifies robustness through endogenous transformations, eliminating dependence on external attack configurations while capturing the intrinsic stability crucial for agent-based systems.

    \item \textbf{Large-scale Empirical Validation and Insights:} We conducted an extensive evaluation of \toolname from three aspects: 

    \begin{itemize}
        \item \textbf{Effectiveness}. Through evaluation of 96 LLMs (0.5B-685B parameters) on MBPP Plus benchmark~\cite{liu2024your}, we reveal that \toolname induces 2.65\%-47.62\% absolute performance drops within ten loops, with robustness patterns that diverge from initial performance rankings. Notably, models like Qwen3-235B-A22B-Instruct-2507 demonstrate superior robustness despite inferior initial capabilities compared to OpenAI's o-series, suggesting that sustained self-consistency may indicate deeper comprehension rather than memorization—offering developers practical guidance for selecting optimal foundation models in iterative, agent-based software systems. To our knowledge, this represents the first comprehensive unified assessment of LLM programming robustness at this scale, encompassing nearly 10 times more models than prior relevant studies and providing the research community with a reliable robustness ranking across the majority of modern LLMs.
        \item \textbf{Reliability}. To ensure that our robustness metrics reflect genuine model capabilities rather than experimental artifacts, we rigorously assessed \toolname's stability against prompt variations and sampling temperatures. Our results demonstrate exceptional consistency, yielding high Spearman rank correlations ($>0.95$) between baseline rankings and those obtained under varying conditions. This confirms that \toolname provides a stable and reproducible evaluation framework, unaffected by minor perturbations in experimental configurations.
        \item \textbf{Extensibility} We demonstrate \toolname's generalization by extending it to code translation tasks, establishing dual transformation loops between different programming languages. 
    This extension supports our framework's general applicability beyond generation-summarization pairs, showcasing its potential for evaluating robustness across diverse software engineering scenarios and proving its easily extensible nature for various dual-task configurations.
    \end{itemize}

    \item \textbf{Open-Source Leaderboard and Resources:} We release a comprehensive online Leaderboard with the results of more than 100 LLMs (Available at: \url{https://evalooop.github.io/}) that enables practitioners to interactively explore model robustness ranking, facilitating informed model selection for automated software engineering workflow. 
    Additionally, we open-source our complete evaluation framework, enabling the research community to reproduce our results and extend robustness evaluation to new models and task configurations, see details by accessing \url{https://github.com/evalooop/EvaLooop}.

\end{itemize}

\section{LLM Robustness in Programming: Related Work and Their Limitations}



The research community has responded by evaluating LLMs' robustness through diverse adversarial techniques~\cite{chen2023evaluating, zhang2024attacks, jha2023codeattack, bavcic2024jy61} and developing corresponding defensive strategies~\cite{improta2025enhancing, yang2024robustness, ge2024demonstration}. However, these evaluation approaches primarily rely on external perturbations strategies. In terms of code-level perturbations, Yefet et al.~\cite{yefet2020adversarial} introduced a gradient-based approach to identify adversarial examples by renaming variables and inserting dead code, while Jha et al.~\cite{jha2023codeattack} proposed CodeAttack to systematically alter code structure through token substitution and style modification. Similarly, Chen et al.~\cite{chen2023evaluating} and Zhang et al.~\cite{zhang2024attacks} explored structure-aware adversarial samples to test model stability against syntactic changes. With the shift toward instruction-tuned models, recent works have focused on natural language perturbations; for instance, Honarvar et al.~\cite{honarvar2025turbulence} proposed Turbulence, a framework that systematically perturbs natural language instructions and question templates to uncover inconsistencies. Despite their contributions, these methods share a common limitation: they rely on externally crafted modifications rather than capturing the model's intrinsic stability.


\noindent\textbf{Limitations of prior works}.
Although adversarial examples expose robustness weakness, we find that different adversarial attacks might ``favor'' different LLMs. As a result, contradictory conclusions could be derived about a model's robustness, which causes model users to struggle in model selection.

To demonstrate the limitations of prior works, we conduct a preliminary experiment by employing three representative attack strategies from a recent work by Sclar et al. \cite{sclar2024quantifying}. 
Each attack perturbs the original prompt from different angles. 
As illustrated in Listing~\ref{lst:adv-exp}, these attacks represent distinct perturbation categories:
\begin{itemize}[wide, labelwidth=!, labelindent=0pt]
    \item \textbf{Case Transformation}: Converting the entire prompt to uppercase to test sensitivity to typographic variations.
    
    \item \textbf{Structural Reformatting}: Reorganizing prompt components with explicit labels and formatting to assess structural dependency.
    
    \item \textbf{Redundant Elaboration}: Injecting verbose instructions and explanatory text to evaluate resistance to information dilution.
\end{itemize}

We systematically apply these three attack methods to measure performance degradation of LLMs by comparing pre- and post-attack pass@1 accuracy.
This approach allows us to quantify each model's vulnerability to different perturbation types while maintaining functional equivalence of the underlying tasks.
As our first attempt, we evaluate 13 state-of-the-art LLMs that have undergone substantial post-training, such as reinforcement learning and instruction-based fine-tuning. 
Our selection encompasses both proprietary closed-source models (e.g., OpenAI's o-series) and open-source alternatives (e.g., Qwen, LLaMA) to ensure comprehensive coverage of contemporary model architectures. 
For each model, we report performance change on MBPP Plus benchmark~\cite{liu2024your} under each attack type, providing a quantitative assessment of adversarial robustness across different perturbation strategies.

\begin{myCodeFrame}{Examples of the three adversarial attacks.}{adv-exp}
\scriptsize
\textbf{Initial prompt:}

Write a function to find the shared elements from the given two lists.
\begin{lstlisting}[language=Python, escapeinside={(*}{*)}, basicstyle=\scriptsize\ttfamily]
assert set(similar_elements((3, 4, 5, 6),(5, 7, 4, 10))) == set((4, 5))
\end{lstlisting}
\vspace{-0.3em}\rule{\linewidth}{0.4pt}\vspace{0.5em}
\textbf{1. Case Transformation:}

\hl{WRITE A FUNCTION TO FIND THE SHARED ELEMENTS FROM THE GIVEN TWO LISTS.}
\begin{lstlisting}[language=Python, escapeinside={(*}{*)}, basicstyle=\scriptsize\ttfamily]
assert set(similar_elements((3, 4, 5, 6),(5, 7, 4, 10))) == set((4, 5))
\end{lstlisting}
\vspace{-0.3em}\rule{\linewidth}{0.4pt}\vspace{0.5em}
\textbf{2. Structural Reformatting:}

\hl{Task:} Write a function to find the shared elements from the given two lists.

\hl{Test Case:}
\begin{lstlisting}[language=Python, escapeinside={(*}{*)}, basicstyle=\scriptsize\ttfamily]
assert set(similar_elements((3, 4, 5, 6),(5, 7, 4, 10))) == set((4, 5))
\end{lstlisting}
\vspace{-0.3em}\rule{\linewidth}{0.4pt}\vspace{0.5em}
\textbf{3. Redundant Elaboration:}

\hl{Please implement the following Python function according to the detailed specification provided below.}

\hl{Function Requirement:} Write a function to find the shared elements from the given two lists.

\hl{Your implementation must satisfy the following test case:}
\begin{lstlisting}[language=Python, escapeinside={(*}{*)}, basicstyle=\scriptsize\ttfamily]
assert set(similar_elements((3, 4, 5, 6),(5, 7, 4, 10))) == set((4, 5))
\end{lstlisting}
\hl{Please ensure your function handles all edge cases and follows Python best practices.}
\end{myCodeFrame}

\begin{table}[t]
\scriptsize
\centering
\caption{Model performance (in terms of pass@1) comparison under adversarial attacks. The lower value indicates a stronger attack and poorer robustness. The strongest attack is in \textbf{bold}.}
\begin{tabular}{lrrrrr}
\toprule
Model     & Original & $ATT_{upper}$  & $ATT_{structure}$ & $ATT_{detail}$ \\
\midrule
Codestral-22B-v0.1             & 0.607 & \textbf{0.648} & 0.706 & 0.669 \\
deepseek-coder-7b-instruct-v1.5 & 0.646 & 0.291 & \textbf{0.011} & 0.638 \\
DeepSeek-Coder-V2-Lite-Instruct & 0.713 & 0.722 & 0.722 & \textbf{0.704} \\
DeepSeek-Coder-V2-Instruct      & 0.769 & 0.767 & \textbf{0.765} & 0.773 \\
Llama-3.1-8B-Instruct           & 0.586 & 0.468 & 0.458 & \textbf{0.360} \\
OpenCoder-8B-Instruct           & 0.703 & \textbf{0.683} & 0.698 & 0.693 \\
CodeQwen1.5-7B-Chat             & 0.698 & \textbf{0.704} & 0.698 & 0.706 \\
Qwen2.5-Coder-32B-Instruct      & 0.762 & 0.759 & 0.751 & \textbf{0.545} \\
Qwen3-235B-A22B-Instruct-2507   & 0.789 & 0.796 & 0.791 & \textbf{0.788} \\
GPT-4.1                         & 0.767 & 0.775 & 0.775 & \textbf{0.743} \\
GPT-4o                          & 0.767 & \textbf{0.754} & 0.767 & 0.759 \\
o3-mini                         & 0.815 & \textbf{0.804} & 0.807 & 0.818 \\
o4-mini                         & 0.818 & 0.818 & 0.810 & \textbf{0.794} \\
\bottomrule
\end{tabular}
\label{tab:model_perf}
\vspace{-4mm}
\end{table}

Table~\ref{tab:model_perf} presents the comparative analysis of model performance under three adversarial attack methods against baseline performance on MBPP Plus. 
We evaluate robustness by measuring pass@1 accuracy change across Case Transformation ($ATT_{upper}$), Structural Reformatting ($ATT_{structure}$), and Redundant Elaboration ($ATT_{detail}$) attacks.

\paragraph{Overall Attack Ineffectiveness} The experimental results reveal that adversarial attacks demonstrate limited effectiveness against modern post-trained LLMs. 
The majority of evaluated models show minimal performance degradation across attack variants. 
For instance, Qwen3-235B-A22B-Instruct-2507 maintain near-identical performance (0.789 → 0.796, 0.791, 0.788), while OpenAI's o-series models (o3-mini, o4-mini) exhibit negligible drops, with some variants even showing slight improvements-a phenomenon consistent with observations by Fang et al.~\cite{fang2025smaller}. 
Similarly, DeepSeek-Coder-V2-Instruct maintains consistent performance (0.769 → 0.767, 0.765, 0.773) across all attack types, indicating robust resilience to external perturbations.

\paragraph{Contradictory Vulnerability Patterns} Among LLMs that do exhibit performance drops, we observe contradictory sensitivity patterns that highlight fundamental limitations in adversarial evaluation. 
deepseek-coder-7b-instruct-v1.5 demonstrates severe vulnerability to Case Transformation (0.646 → 0.291) and catastrophic failure under Structural Reformatting (0.646 → 0.011), yet maintains baseline performance under Redundant Elaboration (0.646 → 0.638). 
Conversely, Qwen2.5-Coder-32B-Instruct shows minimal sensitivity to the first two attacks (0.762 → 0.759, 0.751) but significant degradation under Redundant Elaboration (0.762 → 0.545). 
This contradictory behavior creates an evaluation paradox: which model is more robust than the other? deepseek-coder-7b-instruct-v1.5 performs excellently under verbose prompts but fails catastrophically with structural changes, while Qwen2.5-Coder-32B-Instruct demonstrates the opposite pattern. 
Such inconsistency makes comparative robustness assessment unreliable for model selection, as conclusions heavily rely on the specific attack method employed.


\begin{tcolorbox}[
    colback=boxcolor,
    colframe=bordercolor,
    title={\textbf{Motivating Findings}},
    fonttitle=\bfseries,
    boxrule=1pt,
    arc=3pt,
    left=8pt,
    right=8pt,
    top=8pt,
    bottom=8pt
]
Our analysis highlights two key limits of adversarial attacks:
(1) \textit{Diminished effectiveness} — post-trained LLMs resist external perturbations, making attacks largely ineffective;
(2) \textit{Inherent evaluation bias} — different attack methods favor different models, leading to conflicting robustness results and undermining fair comparison.
\end{tcolorbox}

\section{Proposed Framework: \toolname}
\label{label:app}
\begin{figure}
    \centering
    \centering
    \includegraphics[width=\textwidth]{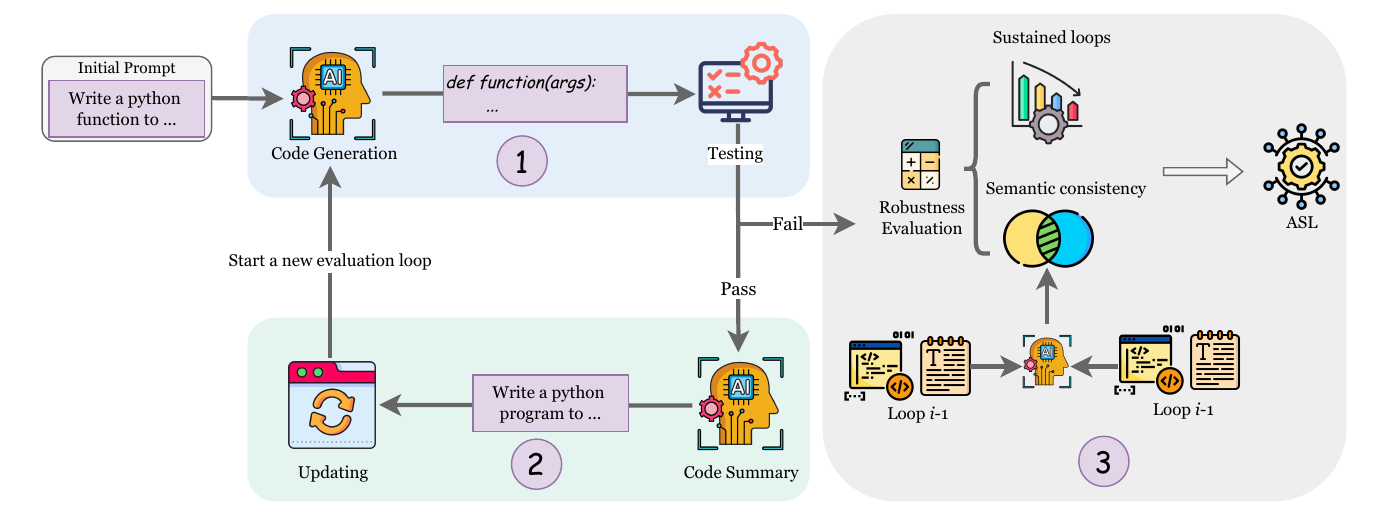}
    \caption{An overview of \toolname implemented by integrating code generation and summarization loop.}
    \label{fig:approach}
\end{figure}

To address the limitations of existing robustness assessment uncovered by our preliminary experiments, we propose a novel framework \toolname for evaluating LLM robustness through self-consistency assessment as presented in Figure~\ref{fig:approach}. 
The core insight driving our approach is that robust LLMs should demonstrate consistent semantic understanding when processing complementary software engineering tasks—such as generating code from natural language specifications and subsequently summarizing that code back into natural language. 
Unlike adversarial approaches that rely on external perturbations, our framework creates endogenous transformations where models iteratively process their own outputs, revealing intrinsic stability patterns crucial in informed model selection for automated software engineering workflow. 
Moreover, this iterative evaluation provides a unified, bias-free metric for comparing LLM stability across diverse architectures and training paradigms.
    
\subsection{Duality Loop}
\label{label:frame}

As illustrated in Figure~\ref{fig:approach}, the evaluation process establishes a cyclical transformation where task outputs are iteratively converted between complementary representations. 
The code generation-summarization pair serves as our exemplar implementation due to its universal applicability and clear semantic relationship: 
code generation demands translating abstract requirements into concrete implementations, 
while code summarization requires extracting and articulating essential logic from existing code. 
However, the core principle of our framework readily extends to other dual task configurations in software engineering, 
such as iterative code translation across multiple programming languages, 
as we demonstrate in Section~\ref{sec:discussion} through a code translation-based extension of \toolname.

Specifically, the evaluation loop operates through the following three key steps:
\circled{1} Starting with an initial natural language prompt specifying a programming task, the target LLM generates executable code that undergoes rigorous testing against predefined test suites. 
\circled{2} Upon successful validation, the same LLM summarizes its generated code back into natural language, creating a new specification for the subsequent iteration. 
This process continues until the generated code fails functional correctness tests.
\circled{3} Upon failed validation, we compute the robustness of the given LLM on the programming task using our proposed Average Sustainable Loops (ASL) metric, which quantifies both loop sustainability and semantic consistency.

The bidirectional nature of this transformation is crucial—it requires the model to maintain semantic fidelity across two distinct cognitive tasks. 
Code generation demands translating abstract requirements into concrete implementations, 
while code summarization requires extracting and articulating the essential logic from existing code. 
A robust LLM should demonstrate consistency across both directions, maintaining the core functionality through successive transformations of its own creation while preserving semantic coherence throughout the iterative evaluation process.

\subsection{Proposed Metric: Average Sustainable Loops (ASL)}
\label{label:asl_metric}

To quantify LLM robustness through our iterative evaluation framework, we introduce the Average Sustainable Loops (ASL) metric. Effective robustness assessment requires capturing two critical dimensions simultaneously: the model's ability to maintain (1) functional performance across increasing loop iterations and (2) the semantic consistency preserved between consecutive transformations.

As illustrated in step 3 of Figure~\ref{fig:approach}, our metric must account for both performance degradation patterns and the quality of semantic preservation throughout the evaluation cycle. Traditional robustness metrics typically focus on binary success/failure outcomes or simple performance drops, failing to distinguish between catastrophic failures due to sudden semantic collapse versus gradual degradation through accumulated drift. 
Our ASL metric addresses this limitation by incorporating semantic similarity analysis that reveals whether functional failures result from significant prompt evolution or sudden robustness breakdown despite minimal semantic changes.
Specifically, we define ASL as a comprehensive measure that incorporates both sustainability duration and semantic consistency:

\begin{equation}
\label{eq:asl}
    \text{ASL} = \frac{\sum_{i=1}^{M} n_i \times i^2 \times s_i}{MT},
\end{equation}

where $M$ represents the maximum number of evaluation loops,
$n_i$ denotes the number of tasks that sustain functionality through exactly $i$ loops,
$T$ represents the total number of tasks in the benchmark dataset,
and $s_i$ represents the average semantic similarity score for tasks failing at loop $i$\footnote{For pure code-to-code transformation tasks (e.g., code translation as discussed in Section~\ref{sec:rq3 design}) where intermediate natural language representations are absent, we set $s_i=1$. In such cases, semantic consistency is strictly determined by binary functional correctness via unit tests, effectively reducing the metric to a weighted measure of sustained loops.}.
The quadratic weighting $i^2$ assigns progressively higher importance to sustained performance across multiple iterations.

For semantic similarity computation, consider a task $\tau$ that sustains $l$ loops before failure. We denote $p_\tau^j$ as the natural language prompt used at loop $j$ for code generation, where $p_\tau^1$ represents the original task specification and $p_\tau^j$ for $j > 1$ represents the code summary generated in the previous loops. The key insight driving our similarity assignment stems from the observation that, from the perspective of LLMs: 
1) two programs that could pass identical test cases should be regarded as semantically equivalent; 
2) following this principle, the prompts used to generate these functionally correct programs should also be considered semantically equivalent, as they specify the same requirements. 
Therefore, for all successful loops, we assign:

\begin{equation}
    s_\tau^{j} = 1 \quad \text{for } j < l \text{ or } l = M.
\end{equation}

For the critical failure boundary where the task terminates at loop $l < M$, we require nuanced evaluation to distinguish between failure due to significant semantic drift versus sudden robustness collapse despite minimal prompt changes:

\begin{equation}
    s_\tau^{l} = \text{Sim}(p_\tau^{l}, p_\tau^{l+1}) \in [0, 1].
\end{equation}

Evaluating semantic similarity at failure boundaries presents significant challenges that standard embedding-based approaches cannot adequately address. 
Traditional similarity metrics rely solely on natural language embeddings, failing to capture the subtle semantic nuances that distinguish functionally correct specifications from those leading to implementation failures. 
The core difficulty lies in requiring comprehensive assessment that considers both the natural language prompt and its corresponding generated code simultaneously—
only through this joint evaluation can we determine whether failure resulted from prompt ambiguity, semantic drift, or sudden model robustness collapse.

\begin{myCodeFrame}{Prompt for LLM-based Semantic Similarity Evaluation.}{similarity-prompt}
\scriptsize
Compare the semantic similarity of these two code generation prompts. Consider both the prompts and their generated code outputs.
\begin{verbatim}

Prompt 1: {prompt1}
Generated Code 1: {code1}

Prompt 2: {prompt2}
Generated Code 2: {code2}

Return only a similarity score between 0 and 1, where:
- 1.0 = semantically equivalent (same intent, requirements, and expected output)
- 0.0 = completely different semantic meaning
- Values between 0 and 1 represent partial semantic overlap

Score:
\end{verbatim}

\end{myCodeFrame}

To address this challenge, we employ an LLM-based evaluator with task-specific prompts designed to perform holistic similarity assessment. 
This approach enables simultaneous consideration of both prompt semantics and code functionality, providing the nuanced evaluation necessary for distinguishing different failure modes at loop boundaries.
The evaluation prompt (shown in Listing~\ref{lst:similarity-prompt}) instructs the evaluator to generate similarity scores between 0 and 1, where higher scores indicate failure occurred despite minimal semantic drift between consecutive prompts, suggesting sudden robustness collapse, 
while lower scores indicate significant semantic divergence at the failure boundary, which may result from either accumulated drift across multiple loops or sudden misinterpretation.

We compute the overall prompt similarity for each task as the per-loop average:

\begin{equation}
\bar{s}_\tau = \frac{1}{l}\sum_{j=1}^{l} s_\tau^{j}.
\end{equation}
\vspace{1mm}

Finally, denoting tasks that sustain exactly $i$ loops as $m_i(1),\dots,m_i(n_i)$, we obtain $s_i$ by averaging over all tasks in this category:

\begin{equation}
    s_i = \frac{1}{n_i}\sum_{k=1}^{n_i} \bar{s}_{\,m_i(k)} = \frac{\sum_{k=1}^{n_i} \sum_{j=1}^{i} s_{m_i(k)}^{j}}{n_i \cdot i}.
\end{equation}
\vspace{1mm}

ASL metric comprehensively addresses the two critical considerations identified earlier. 
First, the quadratic weighting scheme with loop count $i^2$ recognizes that maintaining functionality through multiple iterations demonstrates deeper semantic understanding than initial success alone, implementing a non-linear reward structure where sustained performance across cascading transformations indicates superior robustness. 
Second, the semantic similarity component $s_i$ provides more precise robustness assessment by incorporating the quality of semantic preservation throughout the evaluation process, enabling our metric to account for both the duration of sustained performance and the consistency of model behavior across transformations.

Compared with adversarial attack-based methods that depend on external perturbation strategies, ASL operates through endogenous evaluation where robustness differences reflect genuine model capabilities rather than vulnerability to specific attack methodologies. 
The semantic similarity integration provides fine-grained analysis of failure modes, enabling developers to distinguish between models that gradually degrade through semantic drift versus those experiencing sudden robustness collapse. 
Furthermore, the continuous scoring scale from sustained loop counts naturally accommodates models with varying capability levels, providing informative rankings across the entire performance spectrum rather than binary classifications.

\subsection{Novel Advantages of \toolname}
\label{label:eval loop}

We summarize the following key advantages of \toolname:

\begin{itemize}[wide=0pt]
    \item \textbf{Compared to adversarial attack-based robustness assessment, \toolname is unbiased.} 
        Building upon the duality loop mechanism, \toolname quantifies robustness by measuring how many consecutive transformation cycles a model can sustain before functional crash. 
        This iterative assessment addresses the fundamental limitations of adversarial attack-based evaluation: 
        beyond their diminished effectiveness against post-trained LLMs and inherent bias where different attack strategies favor different models, adversarial approaches rely on external perturbations and fail to capture the internal robustness (i.e., self-consistency) of LLMs—a critical capability for deploying models in complex AI systems where sustained coherence across interactions determines overall reliability.
        
    \item \textbf{\toolname iteratively exposes robustness weakness from the internal of the model.} 
        Unlike many adversarial attacks that rely on externally crafted perturbations, the failing examples in \toolname are generated by the model itself, providing authentic insights into intrinsic model limitations. 
        The iterative framework reveals distinct degradation trajectories that vary significantly across models. Some LLMs demonstrate gradual semantic drift over many loops, while others experience sudden collapse after a few iterations. 
        These patterns reflect fundamental differences in how models represent and maintain semantic information across self-referential transformations. 
        This approach captures degradation emerging from the model's own processing limitations and internal inconsistencies, offering a more genuine assessment of model reliability.

        The sustained loop approach operates independently of external attack configurations. 
        After the initial prompt, all subsequent inputs are generated by the model itself, creating a self-contained evaluation environment. 
        This endogenous design eliminates biases introduced by specific adversarial strategies and enables fair comparison across different model architectures, as robustness differences reflect genuine capabilities rather than vulnerability to particular perturbation methods.
        
    \item \textbf{\toolname is easy to implement, enabling widespread adoption with significant practical impact.} 
    
        Our comprehensive evaluation demonstrates this scalability through assessment of 96 popular LLMs spanning from 0.5B to 685B parameters—substantially exceeding the scope of existing robustness evaluation studies. 
        For comparison, recent adversarial attack studies typically evaluate 5-15 models~\cite{wei2023towards, jha2023codeattack}, while our framework enables evaluation of 5-10 times more models due to its straightforward implementation requirements. 
        This broad applicability facilitates systematic robustness analysis across the entire spectrum of contemporary LLMs.
        To further support practical adoption, we provide an interactive online Leaderboard (Available at: \url{https://evalooop.github.io/}) that enables practitioners to explore model robustness rankings, facilitating informed model selection for various deployment scenarios.

    \item \textbf{\toolname can be easily extended with critical software engineering tasks.}

        Beyond the core generation-summarization cycle, we demonstrate how our framework readily extends to other critical software engineering tasks by integrating different software engineering tasks. Specifically, we implement and evaluate a code translation-based extension where models iteratively translate code across multiple programming languages (e.g., Python → PHP → Ruby → JavaScript → Perl → Python), with each successful translation representing one evaluation loop, as detailed in Section~\ref{sec:discussion}. 
        Our framework's extensibility also enables integration of other dual task configurations such as code and pseudocode conversion~\cite{oda2015learning} and natural language specification and pseudocode transformation~\cite{xu2024core}.
        
    \item \textbf{\toolname offers a customized metric to quantify model robustness accurately.} 
    
        Since our iterative evaluation framework represents a novel approach to robustness assessment, we developed the Average Sustainable Loops (ASL) metric specifically tailored to capture the nuanced robustness patterns revealed through sustained loop evaluation. 
        Unlike traditional binary pass/fail metrics or simple performance drops, ASL incorporates both loop sustainability duration and semantic consistency quality, providing model developers with interpretable scores that directly inform model selection decisions. 
        The metric's design enables clear differentiation between models exhibiting gradual degradation versus sudden collapse, guiding developers toward models best suited for their specific deployment requirements where sustained reliability is paramount.
\end{itemize}

\section{Experimental Setup}


\subsection{Research Questions}
\begin{itemize}[wide, labelwidth=!, labelindent=0pt]
    
    \item \textbf{RQ1 (Effectiveness of the Proposed Framework):} How effectively does \toolname measure LLM robustness in programming?
    
    \item \textbf{RQ2 (Reliability of the Proposed Framework):} How reliable is \toolname to initial prompt variations and temperature settings?

    \item \textbf{RQ3 (Extensibility of the Proposed Framework):} Can we extend \toolname to new variants with other duality tasks, such as code translation?
\end{itemize}

\begin{myCodeFrame}{An example of five variants prompts.}{adv-exp-sum}
\scriptsize
Initial prompt:
Use one sentence to summarize the following code and start with write a python function to:
\begin{verbatim}
```\n{code}\n```
```\nwrite a python function to\n```
\end{verbatim}
\vspace{-0.7em}\rule{\linewidth}{0.4pt}\vspace{0.2em}
Prompt 1 (Generate detailed summarization):
\hl{Summarize what the following code does in natural language:}
\begin{verbatim}
```\n{code}\n```
\end{verbatim}
\hl{Provide a brief explanation in 2-3 sentences. Describe what the code does and its basic inputs/outputs, but avoid technical implementation details and repetitive explanations. Keep it straightforward and concise.}
\vspace{-0.4em}\rule{\linewidth}{0.4pt}\vspace{0.5em}
Prompt 2 (Case transformation):
\hl{USE ONE SENTENCE TO SUMMARIZE THE FOLLOWING CODE AND START WITH WRITE A PYTHON FUNCTION TO:}
\begin{verbatim}
```\n{code}\n```
\end{verbatim}
\verb|```\n|\hl{WRITE A PYTHON FUNCTION TO}\verb|\n```|

\vspace{-0.5em}\rule{\linewidth}{0.4pt}\vspace{0.2em}
Prompt 3 (Structural reformatting):
\hl{Task:} Use one sentence to summarize the following code and start with write a python function to:\\
\hl{1. Input Code:}
\begin{verbatim}
```\n{code}\n```
\end{verbatim}
\hl{2. Output Format:}
\begin{verbatim}
```\nwrite a python function to\n```
\end{verbatim}
\vspace{-0.7em}\rule{\linewidth}{0.4pt}\vspace{0.2em}
Prompt 4 (Simplify):
\hl{Summarize this code in one sentence starting} with `write a python function to':
\begin{verbatim}
```\n{code}\n```
\end{verbatim}
\hl{Summary:} write a python function to

\vspace{-0.1em}\rule{\linewidth}{0.4pt}\vspace{0.2em}
Prompt 5 (Redundant elaboration):
\hl{Please analyze the provided Python code below and generate a concise one-sentence summary. Your response must begin with the exact phrase} `write a python function to'.
\hl{Code to analyze:}
\begin{verbatim}
```python\n{code}\n```
\end{verbatim}
\hl{Please provide your summary in the following format:}
\begin{verbatim}
```\nwrite a python function to\n```
\end{verbatim}
\end{myCodeFrame}

\subsection{Methodology for Answering RQ1}

This investigation aims to validate \toolname's ability to differentiate robustness levels and reveal stability patterns that conventional metrics cannot capture.
In detail, we evaluate 96 LLMs spanning parameter scales from 0.5B to 685B, covering the majority of mainstream models currently available, including both open-source and close-source. 
Our evaluation employs MBPP Plus~\cite{liu2024your}, an enhanced version of the original MBPP benchmark~\cite{austin2021program} that applies rigorous quality filtering to programming tasks and expands the test case corpus by 35-fold. 
This enhanced benchmark provides comprehensive evaluation coverage specifically suited for our robustness assessment framework.
To maintain consistency with the benchmark's code generation prompts, we design complementary prompts for the code summarization task, as shown the initial prompt in Listing~\ref{lst:adv-exp-sum}. 
It instruct LLMs to generate code summaries that follow the same format as the initial generation prompts, ensuring semantic alignment across the dual-task evaluation loop.
We analyze robustness rankings using our proposed ASL metric to identify LLMs that demonstrate superior self-consistency independent of initial performance. 
This includes examining cases where models with relatively lower initial performance exhibit higher robustness scores, potentially indicating deeper semantic understanding versus memorization patterns. 
The comprehensive scale of our evaluation enables robust statistical analysis of robustness patterns across model families, parameter scales, and training approaches.


\subsection{Methodology for Answering RQ2}
To assess the reliability of \toolname under varying experimental conditions, we investigate two critical factors that could influence ASL measurements: prompt variations and temperature settings. 
This analysis ensures our framework provides stable robustness assessments across different deployment configurations.

\noindent\textbf{Prompt Reliability Analysis} 
We assess the reliability of \toolname to prompt variations by focusing on the code summarization component, which drives the iterative transformation loops in our evaluation. 
Different prompt formulations are applied to every loop iteration to determine whether observed robustness differences reflect genuine model capabilities rather than prompt-specific artifacts. 
As shown in Listing~\ref{lst:adv-exp-sum}, we employ five distinct prompt variants: the three adversarial strategies from Methodology for RQ1 (Case Transformation, Structural Reformatting, and Redundant Elaboration) plus two additional approaches—generating detailed summarization and prompt simplification. 
Unlike RQ1, which examines model vulnerability to these perturbations, RQ3 investigates whether \toolname produces consistent robustness rankings despite input variations across the summarization phase. 
A robust evaluation framework should demonstrate minimal ranking perturbations across semantically equivalent prompts, indicating that observed robustness differences reflect genuine model capabilities rather than prompt-specific artifacts.

\noindent\textbf{Temperature Reliability Analysis} We systematically evaluate ASL score variations across different sampling temperatures to understand the framework's behavior under varying generation stochasticity. 
We test four additional temperature settings: 0.2, 0.4, 0.6, and 0.8, comparing them against the default temperature configuration.

In this RQ, we experiment on the same 13 LLMs from RQ1, ensuring consistency with our adversarial attack comparison while maintaining experimental feasibility. 
To quantify the consistency of model rankings across different prompts, we calculate the \textit{Spearman rank correlation coefficient} between the baseline rankings and the average prompt and temperature rankings. 
A high correlation coefficient generally indicates a strong positive correlation, demonstrating that the relative performance of LLMs remains highly stable across different prompt variations and temperature settings.

\subsection{Methodology for Answering RQ3}
\label{sec:rq3 design}

To demonstrate the generalizability and practical utility of our framework, we extend \toolname beyond its base configuration. While the standard \textbf{generation-summarization loop} serves as a general-purpose probe for assessing an LLM's fundamental programming capabilities—specifically its ability to translate abstract requirements and articulate complex logic—we introduce a \textbf{code translation variant} to simulate real-world software evolution scenarios. This extension is particularly relevant for large-scale \textbf{project migration} tasks, such as porting legacy systems from Java to modern, memory-safe languages like Rust, where maintaining semantic equivalence across different programming paradigms is paramount.

To measure the \toolname in this configuration, we update the ASL metric by the definition of the ASL metric (Section~\ref{label:asl_metric}) accordingly. Since the translation chain operates entirely within the code domain without intermediate natural language summaries, we \textbf{omit the semantic similarity component} ($s_{\tau}$) to prioritize a purely functional evaluation. The robustness is thus quantified by the sustained number of functional iterations where the code passes language-specific test suites. This provides a more rigorous and deterministic measure of a model's structural endurance across diverse programming environments. The code translation loop establishes a continuous chain where each successful conversion represents one iteration. Starting with correct source code, the LLM receives prompts to translate the implementation into a subsequent language (e.g., Python $\rightarrow$ PHP), followed by rigorous validation against predefined.

\subsection{Implementation Details}

\begin{figure*}[th]
    \centering
    \includegraphics[width=\linewidth]{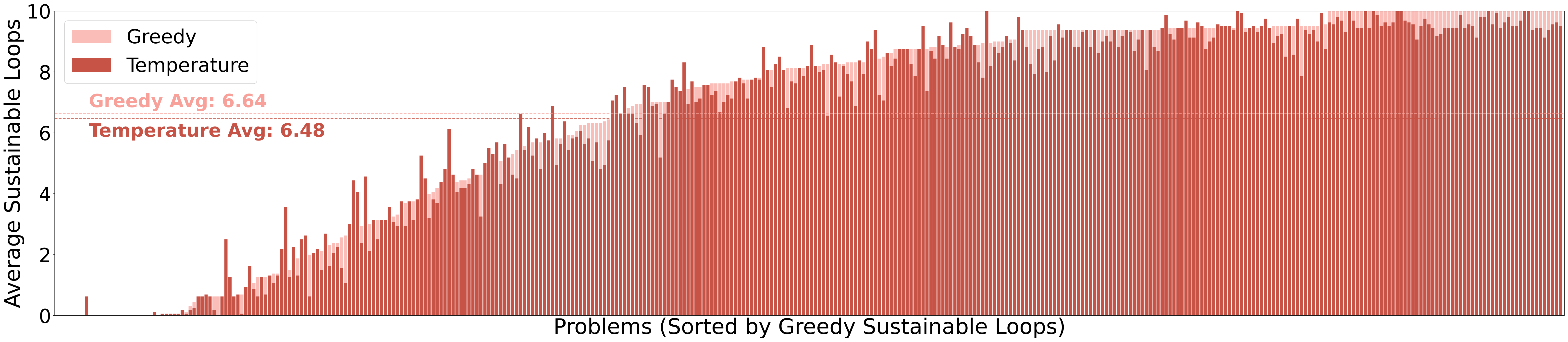}
    \caption{Distribution of average sustainable loops per task in MBPP Plus.}
    \label{lab:decoding}
\end{figure*}

\noindent\textbf{Greedy Decoding vs. Temperature Sampling}
Figure~\ref{lab:decoding} presents the performance distribution of average sustainable loops per task in the MBPP Plus benchmark across 13 LLMs (same as RQ1 and RQ3) using either greedy decoding or temperature-controlled stochastic sampling.
Our comprehensive evaluation, conducted through five times experimental runs and averaged results, reveals an intriguing finding: both decoding strategies achieve nearly identical performance, with greedy decoding averaging 6.38 sustainable loops and temperature sampling averaging 6.37 sustainable loops, which is similar to previous work~\cite{liu2024your}.
This convergence in performance is particularly noteworthy given that temperature sampling is widely adopted as the default decoding mechanism in many production LLMs (e.g., ChatGPT employs a default temperature of 0.7), primarily valued for its capacity to introduce variability and enhance human-like expression diversity.
However, our results suggest that in the context of code generation tasks, the stochastic exploration benefits of temperature sampling do not translate to measurable performance improvements.
This might be becasue code generation inherently demands strict adherence to precise semantic and syntactic rules, creating a more constrained solution space where the highest-probability tokens selected by greedy decoding are often optimal.
Besides, the deterministic nature of greedy decoding may actually be advantageous in programming tasks, where consistency and logical coherence are paramount over linguistic creativity.
Given the equivalent performance and the superior reproducibility afforded by greedy decoding, we adopt greedy decoding as our primary evaluation strategy for all experiments of our research questions.

\noindent\textbf{Hardware environment and hyperparameter settings}.
We evaluate proprietary LLMs through their official APIs (OpenAI, Anthropic, etc.) and deploy open-source LLMs locally using VLLM~\cite{kwon2023efficient} across eight NVIDIA H200 GPUs. All experiments use consistent hyperparameters: 
temperature=0.0 and top-p=1.0 for deterministic generation (greedy decoding), maximum 1024 tokens for code generation and summarization. Besides, we set the maximum number of evaluation loop (\ie $M$) as 10 and use gpt-4-turbo-2024-04-09 for similarity evaluator.

\section{Results}

\subsection{Answer to RQ1: Framework Effectiveness}

\begin{table*}[t]
\centering
\scriptsize
\setlength{\tabcolsep}{3pt}
\caption{Robustness ranking benchmarked by \toolname. Numbers in parentheses denote the change in ASL ranking relative to the PassRate ranking.}
\label{tab:asl_results}
\begin{minipage}[t]{0.49\textwidth}
\centering
\begin{tabularx}{\linewidth}{r >{\raggedright\arraybackslash}X c}
\toprule
\textbf{ } & \textbf{Model} & \textbf{ASL} \\
\midrule
1  & Claude Opus 4 & 7.786 ($\darkgreen\Uparrow$\ \ 2) \\
2  & Qwen3-235B-A22B-Instruct-2507 & 7.689 ($\darkgreen\Uparrow$\ \ 6) \\
3  & Qwen3-Coder-480B-A35B-Instruct & 7.599 ($\darkgreen\Uparrow$\ \ 6) \\
4  & Claude Sonnet 4 & 7.518 ($\darkgreen\Uparrow$\ \ 6) \\
5  & DeepSeek-V3 & 7.481 ($\darkgreen\Uparrow$\ \ 1) \\
6  & o3-mini & 7.456 ($\darkred\Downarrow$\ \ 2) \\
7  & o3 & 7.383 (\ \ =\ \ ) \\
8  & GPT-4.1 & 7.355 ($\darkgreen\Uparrow$\ \ 4) \\
9  & o4-mini & 7.317 ($\darkred\Downarrow$\ \ 8) \\
10 & GPT-4.1-mini & 7.290 ($\darkgreen\Uparrow$\ \ 1) \\
11 & o1 & 7.285 ($\darkred\Downarrow$\ \ 9) \\
12 & o1-mini & 7.254 ($\darkred\Downarrow$\ \ 7) \\
13 & Qwen2.5-Coder-32B-Instruct & 7.253 ($\darkgreen\Uparrow$\ \ 6) \\
14 & Gemini 2.5 Flash-Lite & 7.178 ($\darkgreen\Uparrow$\ \ 1) \\
15 & Qwen3-30B-A3B-Instruct-2507 & 7.124 ($\darkgreen\Uparrow$\ 11) \\
16 & Claude Haiku 3.5 & 7.097 ($\darkgreen\Uparrow$\ \ 1) \\
17 & DeepSeek-Coder-V2-Instruct & 7.070 ($\darkgreen\Uparrow$\ \ 3) \\
18 & Gemini 2.0 Flash & 7.041 ($\darkred\Downarrow$\ \ 2) \\
19 & GPT-4o & 7.039 ($\darkgreen\Uparrow$\ \ 4) \\
20 & DeepSeek-V2.5 & 6.945 ($\darkgreen\Uparrow$\ \ 1) \\
21 & Llama-4-Maverick-17B-128E-Instruct & 6.922 ($\darkred\Downarrow$\ \ 8) \\
22 & Gemini 2.5 Flash & 6.922 ($\darkred\Downarrow$\ \ 4) \\
23 & gemma-3-27b-it & 6.905 ($\darkgreen\Uparrow$\ \ 4) \\
24 & Gemini 2.5 Pro & 6.842 ($\darkred\Downarrow$\ 10) \\
25 & gemma-3-12b-it & 6.839 ($\darkgreen\Uparrow$\ 12) \\
26 & GPT-4-Turbo & 6.820 ($\darkgreen\Uparrow$\ \ 2) \\
27 & GLM-4.5-Air & 6.809 ($\darkgreen\Uparrow$\ \ 6) \\
28 & Llama-3.1-405B-Instruct & 6.787 ($\darkgreen\Uparrow$\ 10) \\
29 & Gemini 2.0 Flash-Lite & 6.775 (\ \ =\ \ ) \\
30 & Qwen2.5-Coder-14B-Instruct & 6.676 (\ \ =\ \ ) \\
31 & Qwen3-Coder-30B-A3B-Instruct & 6.669 ($\darkred\Downarrow$\ \ 9) \\
32 & NextCoder-32B & 6.610 ($\darkred\Downarrow$\ \ 8) \\
33 & NextCoder-14B & 6.520 ($\darkgreen\Uparrow$\ \ 1) \\
34 & Llama-3.3-70B-Instruct & 6.517 ($\darkgreen\Uparrow$\ 10) \\
35 & OpenCoder-8B-Instruct & 6.490 ($\darkgreen\Uparrow$\ 12) \\
36 & Seed-Coder-8B-Instruct & 6.452 ($\darkred\Downarrow$\ \ 5) \\
37 & Mistral-Large-Instruct-2407 & 6.422 ($\darkgreen\Uparrow$\ \ 8) \\
38 & CodeQwen1.5-7B-Chat & 6.410 ($\darkgreen\Uparrow$\ 11) \\
39 & DeepSeek-Coder-V2-Lite-Instruct & 6.377 ($\darkgreen\Uparrow$\ \ 7) \\
40 & Qwen2.5-Coder-7B-Instruct & 6.340 (\ \ =\ \ ) \\
41 & Hermes-3-Llama-3.1-405B & 6.295 (\ \ =\ \ ) \\
42 & gemma-3n-E4B-it & 6.279 ($\darkred\Downarrow$\ \ 3) \\
43 & GPT-3.5-Turbo & 6.257 ($\darkgreen\Uparrow$\ \ 5) \\
44 & GLM-4.5 & 6.226 ($\darkred\Downarrow$\ 19) \\
45 & GLM-4-9B-0414 & 6.198 ($\darkgreen\Uparrow$\ \ 8) \\
46 & Ling-lite-1.5 & 6.177 ($\darkred\Downarrow$\ \ 4) \\
47 & Pixtral-Large-Instruct-2411 & 6.150 ($\darkgreen\Uparrow$\ \ 9) \\
48 & MiniCPM4-8B & 6.137 ($\darkred\Downarrow$\ 13) \\

\bottomrule
\end{tabularx}
\end{minipage}\hfill
\begin{minipage}[t]{0.49\textwidth}
\centering
\begin{tabularx}{\linewidth}{r >{\raggedright\arraybackslash}X c}
\toprule
\textbf{} & \textbf{Model} & \textbf{ASL} \\
\midrule

49 & Mistral-Small-3.2-24B-Instruct-2506 & 6.029 ($\darkgreen\Uparrow$\ \ 1) \\
50 & Llama-3.1-70B-Instruct & 6.023 ($\darkred\Downarrow$\ \ 7) \\
51 & Meta-Llama-3-70B-Instruct & 6.017 ($\darkgreen\Uparrow$\ \ 9) \\
52 & Qwen3-4B-Instruct-2507 & 5.986 ($\darkgreen\Uparrow$\ \ 7) \\
53 & Hermes-3-Llama-3.1-70B & 5.983 ($\darkgreen\Uparrow$\ \ 1) \\
54 & Qwen2.5-Coder-3B-Instruct & 5.946 ($\darkgreen\Uparrow$\ \ 7) \\
55 & Mistral-Small-3.1-24B-Instruct-2503 & 5.937 ($\darkgreen\Uparrow$\ \ 2) \\
56 & DeepSeek-V2-Chat & 5.854 ($\darkred\Downarrow$\ \ 4) \\
57 & Baichuan-M2-32B & 5.827 ($\darkred\Downarrow$\ 25) \\
58 & deepseek-coder-33b-instruct & 5.767 ($\darkred\Downarrow$\ \ 7) \\
59 & gemma-3-4b-it & 5.766 ($\darkgreen\Uparrow$\ \ 4) \\
60 & Mistral-Small-Instruct-2409 & 5.700 ($\darkgreen\Uparrow$\ \ 7) \\
61 & Mixtral-8x22B-Instruct-v0.1 & 5.575 ($\darkgreen\Uparrow$\ \ 5) \\
62 & Codestral-22B-v0.1 & 5.443 ($\darkgreen\Uparrow$\ 10) \\
63 & deepseek-coder-7b-instruct-v1.5 & 5.423 ($\darkgreen\Uparrow$\ \ 1) \\
64 & Qwen2.5-Coder-1.5B-Instruct & 5.319 ($\darkgreen\Uparrow$\ 11) \\
65 & Llama-4-Scout-17B-16E-Instruct & 5.224 ($\darkred\Downarrow$\ 29) \\
66 & gemma-3n-E2B-it & 5.051 ($\darkred\Downarrow$\ \ 4) \\
67 & Phi-4-reasoning & 4.897 ($\darkred\Downarrow$\ 12) \\
68 & Ministral-8B-Instruct-2410 & 4.844 ($\darkgreen\Uparrow$\ 11) \\
69 & LFM-7B & 4.743 ($\darkgreen\Uparrow$\ 11) \\
70 & Phi-3.5-MoE-instruct & 4.713 ($\darkgreen\Uparrow$\ \ 1) \\
71 & Phi-3.5-mini-instruct & 4.706 ($\darkgreen\Uparrow$\ 12) \\
72 & Mixtral-8x7B-Instruct-v0.1 & 4.687 ($\darkgreen\Uparrow$\ \ 6) \\
73 & Ling-plus & 4.521 ($\darkgreen\Uparrow$\ \ 8) \\
74 & Hermes-3-Llama-3.1-8B & 4.446 ($\darkgreen\Uparrow$\ \ 8) \\
75 & Meta-Llama-3-8B-Instruct & 4.310 ($\darkgreen\Uparrow$\ \ 2) \\
76 & Phi-4-mini-instruct & 4.008 ($\darkgreen\Uparrow$\ 10) \\
77 & Qwen3-235B-A22B & 3.933 ($\darkred\Downarrow$\ \ 9) \\
78 & Qwen2.5-Coder-0.5B-Instruct & 3.424 ($\darkgreen\Uparrow$\ 11) \\
79 & DeepSeek-V2-Lite-Chat & 3.280 ($\darkgreen\Uparrow$\ \ 5) \\
80 & Moonlight-16B-A3B-Instruct & 3.181 ($\darkred\Downarrow$\ 11) \\
81 & Mistral-7B-Instruct-v0.3 & 3.129 ($\darkgreen\Uparrow$\ \ 7) \\
82 & NextCoder-7B & 3.123 ($\darkred\Downarrow$\ 24) \\
83 & Seed-Coder-8B-Reasoning & 3.106 ($\darkgreen\Uparrow$\ \ 4) \\
84 & Phi-3-mini-4k-instruct & 3.012 ($\darkred\Downarrow$\ 11) \\
85 & Mistral-7B-Instruct-v0.2 & 2.895 ($\darkgreen\Uparrow$\ \ 6) \\
86 & Llama-3.1-8B-Instruct & 2.640 ($\darkred\Downarrow$\ 10) \\
87 & DeepSeek-R1-Distill-Llama-70B & 2.373 ($\darkred\Downarrow$\ 17) \\
88 & Phi-3-medium-4k-instruct & 2.368 ($\darkgreen\Uparrow$\ \ 4) \\
89 & OpenCoder-1.5B-Instruct & 2.352 ($\darkred\Downarrow$\ 15) \\
90 & Qwen3-4B-Thinking-2507 & 2.092 (\ \ =\ \ ) \\
91 & GLM-4-32B-0414 & 1.959 ($\darkred\Downarrow$\ 26) \\
92 & Jan-v1-4B & 1.847 ($\darkred\Downarrow$\ \ 7) \\
93 & Qwen3-1.7B & 1.636 (\ \ =\ \ ) \\
94 & Claude Haiku 3 & 1.571 ($\darkgreen\Uparrow$\ \ 1) \\
95 & deepseek-coder-1.3b-instruct & 1.340 ($\darkred\Downarrow$\ \ 1) \\
96 & deepseek-coder-6.7b-instruct & 1.107 (\ \ =\ \ ) \\

\bottomrule
\end{tabularx}
\end{minipage}
\vspace{-5mm}
\end{table*}

Table~\ref{tab:asl_results} presents the comprehensive robustness evaluation of 96 LLMs using \toolname, ranked by their ASL scores from highest to lowest. 
Models with high ASL scores demonstrate both the ability to maintain functional correctness across extended iterations and semantic coherence throughout self-referential transformations.
The numbers in parentheses indicate the change in ASL ranking relative to each model's pass@1 accuracy ranking, revealing how robustness assessment differs from conventional performance metrics.

\noindent\textbf{Robustness vs. Initial Performance Divergence} 
The results demonstrate substantial divergence between robustness rankings and initial performance capabilities. 
The majority of models exhibit significant ranking changes when evaluated through the robustness lens: 
88 out of 96 models (91.7\%) show ranking shifts, 
with a mean absolute deviation of 6.9 positions. 
Over one-quarter of models (27.1\%) experience large shifts of 10 or more positions, 
and nearly one in ten model pairs have their relative ordering reversed between PassRate and ASL evaluations. 
Notable examples include Qwen3-235B-A22B-Instruct-2507, which rises 6 positions in robustness ranking despite already strong initial performance, and conversely, o1 and o4-mini, which drop 9 and 8 positions respectively, indicating that high initial performance does not guarantee sustained stability through iterative transformations.
Besides, there are some models experiencing significant drops (e.g., Llama-4-Scout-17B-16E-Instruct dropping 29 positions, GLM-4-32B-0414 dropping 26 positions), which reveals brittleness that becomes apparent only through sustained iterative evaluation. 
Conversely, models with positive ranking changes demonstrate robustness capabilities that exceed their initial performance suggestions, offering valuable insights for model selection in deployment scenarios requiring sustained reliability.

The ASL scores reveal pronounced differences in intrinsic stability across models, with scores ranging from 7.786 (Claude Opus 4) to 1.107 (deepseek-coder-6.7b-instruct). 
This wide distribution confirms that robustness represents a distinct capability dimension independent of initial coding competence, validating the necessity of specialized robustness evaluation frameworks. 
Remarkably, several models demonstrate superior robustness despite lower initial performance rankings—for instance, gemma-3-12b-it rises 12 positions, and OpenCoder-8B-Instruct gains 12 positions, suggesting that sustained self-consistency may indicate deeper comprehension mechanisms rather than mere memorization patterns.

\noindent\textbf{Impact of Model Scale on Robustness.}
The extensive parameter range of our evaluated models (0.5B to 685B) allows for a granular investigation into the relationship between model scale and intrinsic robustness. Our analysis reveals a complex landscape that challenges the linear "Scaling Law" assumption.

\noindent\textit{(1) Divergence at the Ultra-Large Scale.} While top-tier performance is generally dominated by large models (e.g., \textit{Claude Opus 4} at Rank 1 and \textit{Qwen3-235B-A22B-Instruct-2507} at Rank 2), mere parameter magnitude does not guarantee superior self-consistency.
Counter-intuitively, several ultra-large-scale models exhibit "Robustness Inefficiency."
For instance, \textit{Llama-3.1-405B-Instruct}, despite its massive size, ranks only 28th (ASL 6.787), significantly under-performing compared to the much smaller \textit{Qwen2.5-Coder-32B-Instruct}.
Similarly, \textit{GLM-4.5} (355B total parameters) ranks 44th, showing a steep drop of 19 positions relative to its initial Pass@1 capability.
Even community-finetuned variants like \textit{Hermes-3-Llama-3.1-405B} (Rank 41) fail to rectify this instability.
This suggests that beyond a certain threshold, increasing parameter count may yield diminishing returns for self-consistency, potentially due to the increased difficulty in maintaining semantic coherence across a vaster knowledge space without specialized alignment.

\noindent\textit{(2) The Rise of ``Mini" Giants.}
In stark contrast to the under-performing giants, we observe a ``Small-Model Resilience" phenomenon.
Highly optimized compact models, such as \textit{o3-mini} (Rank 6), \textit{GPT-4o-mini} (Rank 10), and Qwen2.5-Coder-32B-Instruct (Rank 13), achieve top-tier robustness scores (ASL $>$ 7.25), outperforming \textit{Llama-3.1-405B} and \textit{Mixtral-8x22B} (Rank 61).
This indicates that advanced post-training techniques (e.g., reinforcement learning~\cite{shao2024deepseekmath, ouyang2022training}) are more critical for sustained robustness than raw parameter count.

\noindent\textit{(3) Open vs. Closed Source Dynamics.}
When comparing model ecosystems, closed-source models generally exhibit higher parameter efficiency. 
For example, \textit{GPT-4o} surpasses the majority of open-source models regardless of size.
However, the gap is narrowing. \textit{Qwen3-235B} (Open) effectively rivals \textit{Claude Opus 4} (Closed), 
and \textit{DeepSeek-V3} (Rank 5) demonstrates that open-weights models can achieve "closed-tier" robustness without relying on the massive computational cost of dense Llama-405B series.
These comparisons collectively imply that for deploying reliable agent systems, developers should prioritize models with proven self-consistency records (e.g., Qwen or o-series) rather than blindly selecting the largest available model checkpoint.

\begin{tcolorbox}[
    colback=boxcolor,
    colframe=bordercolor,
    title={\textbf{Answer Summary to RQ1}},
    fonttitle=\bfseries,
    boxrule=1pt,
    arc=3pt,
    left=8pt,
    right=8pt,
    top=8pt,
    bottom=8pt
]
Our large-scale assessment on 96 LLMs reveals:
(1) \textit{Robustness-performance divergence:} 88 out of 96 models show significant ranking changes (ranging from -29 to +12) when evaluated for sustained stability versus initial performance;
(2) \textit{Scaling Law Deviations:} Robustness does not linearly scale with parameter count. We observe ``Robustness Inefficiency'' in ultra-large dense models (e.g., Llama-3.1-405B) and ``Small-Model Resilience'' in compact models (e.g., o3-mini), demonstrating that architectural efficiency and post-training alignment are more critical determinants of self-consistency than massive scale.
\end{tcolorbox}

\subsection{Answer to RQ2: Framework Reliability}
\begin{table}[ht]
\centering
\caption{Prompt sensitivity results. Numbers in parentheses are the rank within the current column; the final averaged prompt rank is the mean of the five prompt-specific ranks (Prompt1--Prompt5).}
\begin{adjustbox}{max width=\linewidth}
\begin{tabular}{lccccccc}
\hline
\textbf{Model} & \textbf{Baseline} & \textbf{Prompt1} & \textbf{Prompt2} & \textbf{Prompt3} & \textbf{Prompt4} & \textbf{Prompt5} & \textbf{Avg. Prompt Rank} \\
\hline
Qwen3-235B-A22B-Instruct-2507 & 7.689 (1) & 7.654 (1) & 7.603 (1) & 7.655 (1) & 7.710 (1) & 7.553 (1) & \textbf{1.00} \\
o3-mini & 7.456 (2) & 7.272 (4) & 7.489 (2) & 7.481 (2) & 7.523 (2) & 7.355 (2) & \textbf{2.40} \\
GPT-4.1 & 7.355 (3) & 7.475 (3) & 7.372 (3) & 7.431 (3) & 7.445 (4) & 7.348 (3) & \textbf{3.20} \\
o4-mini & 7.317 (4) & 7.482 (2) & 7.299 (4) & 7.367 (4) & 7.447 (3) & 7.335 (4) & 3.40 \\
Qwen2.5-Coder-32B-Instruct & 7.253 (5) & 7.029 (7) & 7.022 (7) & 7.087 (6) & 7.003 (7) & 7.020 (7) & 6.80 \\
DeepSeek-Coder-V2-Instruct & 7.070 (6) & 7.250 (5) & 7.147 (6) & 6.908 (7) & 7.054 (6) & 7.081 (6) & 6.00 \\
GPT-4o & 7.039 (7) & 7.231 (6) & 7.207 (5) & 7.163 (5) & 7.148 (5) & 7.131 (5) & 5.20 \\
OpenCoder-8B-Instruct & 6.490 (8) & 6.236 (10) & 6.749 (8) & 6.776 (8) & 2.913 (12) & 6.085 (10) & 9.60 \\
DeepSeek-Coder-V2-Lite-Instruct & 6.377 (9) & 6.597 (9) & 6.234 (10) & 6.293 (9) & 5.949 (9) & 6.094 (9) & 9.20 \\
CodeQwen1.5-7B-Chat & 6.340 (10) & 6.753 (8) & 6.277 (9) & 6.198 (10) & 6.173 (8) & 6.146 (8) & 8.60 \\
Codestral-22B-v0.1 & 5.443 (11) & 5.530 (12) & 5.209 (12) & 5.221 (11) & 5.404 (10) & 5.272 (12) & 11.40 \\
deepseek-coder-7b-instruct-v1.5 & 5.423 (12) & 5.804 (11) & 5.572 (11) & 2.428 (12) & 4.969 (11) & 5.580 (11) & 11.20 \\
Llama-3.1-8B-Instruct & 2.640 (13) & 2.683 (13) & 3.469 (13) & 2.065 (13) & 1.583 (13) & 5.189 (13) & 13.00 \\
\hline
\end{tabular}
\end{adjustbox}
\vspace{-3mm}
\label{tab:prompt_ranks}
\end{table}

\begin{table}[ht]
\centering
\caption{Temperature sensitivity results. Numbers in parentheses are the rank within the current column; the final averaged temperature rank is the mean of the four temperature-specific ranks (0.2, 0.4, 0.6, and 0.8).}
\begin{adjustbox}{max width=\linewidth}
\begin{tabular}{lcccccc}
\hline
\textbf{Model} & \textbf{Baseline} & \textbf{0.2} & \textbf{0.4} & \textbf{0.6} & \textbf{0.8} & \textbf{Avg. Temperature Rank} \\
\hline
Qwen3-235B-A22B-Instruct-2507 & 7.689 (1) & 7.643 (1) & 7.673 (1) & 7.662 (1) & 7.686 (1) & 1.00 \\
o3-mini & 7.456 (2) & 7.375 (3) & 7.458 (2) & 7.410 (2) & 7.389 (3) & 2.50 \\
GPT-4.1 & 7.355 (3) & 7.209 (4) & 7.315 (4) & 7.389 (4) & 7.214 (4) & 4.00 \\
o4-mini & 7.317 (4) & 7.622 (2) & 7.389 (3) & 7.395 (3) & 7.431 (2) & 2.50 \\
Qwen2.5-Coder-32B-Instruct & 7.253 (5) & 7.140 (6) & 7.082 (6) & 7.158 (5) & 7.067 (6) & 5.75 \\
DeepSeek-Coder-V2-Instruct & 7.070 (6) & 7.186 (5) & 7.124 (5) & 7.141 (6) & 7.151 (5) & 5.25 \\
GPT-4o & 7.039 (7) & 6.961 (7) & 7.036 (7) & 6.783 (7) & 6.611 (7) & 7.00 \\
OpenCoder-8B-Instruct & 6.490 (8) & 5.838 (10) & 5.827 (9) & 5.924 (9) & 5.625 (9) & 9.25 \\
DeepSeek-Coder-V2-Lite-Instruct & 6.377 (9) & 6.419 (8) & 6.385 (8) & 6.183 (8) & 6.273 (8) & 8.00 \\
CodeQwen1.5-7B-Chat & 6.340 (10) & 6.334 (9) & 5.381 (11) & 4.916 (12) & 4.784 (10) & 10.50 \\
Codestral-22B-v0.1 & 5.443 (11) & 5.207 (12) & 5.025 (12) & 4.959 (11) & 4.734 (11) & 11.50 \\
deepseek-coder-7b-instruct-v1.5 & 5.423 (12) & 5.438 (11) & 5.467 (10) & 5.392 (10) & 4.701 (12) & 10.75 \\
Llama-3.1-8B-Instruct & 2.640 (13) & 2.804 (13) & 2.525 (13) & 1.985 (13) & 1.246 (13) & 13.00 \\
\hline
\end{tabular}
\end{adjustbox}
\vspace{-5mm}
\label{tab:avg_ranks}
\end{table}

Tables~\ref{tab:prompt_ranks} and~\ref{tab:avg_ranks} present the reliability analysis of \toolname across prompt variations and different temperature settings. 
We evaluate our framework stability by examining the variation of standard ASL score and ranking consistency across different experimental conditions, with rankings shown in parentheses for each model.

\noindent\textbf{Prompt Reliability Analysis} To assess prompt reliability, we compute the ranking for each LLM across all prompt variants (Prompts 1-5) and compare it with the original rankings. 
Our analysis reveals remarkable stability in model rankings despite prompt variations. 
For instance, Qwen3-235B-A22B-Instruct-2507 maintains consistent top ranking (rank 1) across all prompt conditions, 
while models like Llama-3.1-8B-Instruct consistently occupy the bottom position (rank 13).
Computing the average rankings across prompt variants yields: Qwen3 (1.0), o3-mini (2.4), gpt-4.1 (3.2), o4-mini (3.4), demonstrating minimal deviation from baseline rankings. 
Moreover, the Spearman rank correlation coefficient between original and average prompt rankings is 0.951, indicating exceptionally high framework stability against prompt modifications.

\noindent\textbf{Temperature Reliability Analysis} Our temperature reliability analysis also reveals similar stability patterns across sampling configurations. 
Computing average rankings across temperatures 0.2-0.8: 
Qwen3 (1.0), o3-mini (2.5), gpt-4.1 (4.0), DeepSeek-Coder-V2-Instruct (5.25), Qwen2.5-Coder-32B (5.75), show minimal ranking perturbations from baseline. 
The Spearman rank correlation coefficient between original and average temperature rankings is \textbf{0.974}, demonstrating even higher stability than prompt variations.
Notably, higher temperatures generally correlate with slight ASL score reductions (e.g., Codestral-22B-v0.1: 5.443 → 4.734, CodeQwen1.5-7B-Chat: 6.340 → 4.784 at temperature 0.8), reflecting increased generation stochasticity. 
However, relative model rankings remain largely preserved despite these score variations.


\begin{tcolorbox}[
    colback=boxcolor,
    colframe=bordercolor,
    title={\textbf{Answer Summary to RQ2}},
    fonttitle=\bfseries,
    boxrule=1pt,
    arc=3pt,
    left=8pt,
    right=8pt,
    top=8pt,
    bottom=8pt
]
Our reliability analysis reveals: 
(1) \textit{Prompt robustness} - Spearman correlation of 0.951 between baseline and prompt-varied rankings, with minimal ranking perturbations (±1-2 positions) across semantic-preserving modifications; 
(2) \textit{Temperature stability} - Even higher correlation of 0.974 across temperature settings (0.2-0.8), indicating that observed robustness differences reflect genuine model capabilities rather than experimental artifacts, validating \toolname as a reliable evaluation framework for fair model comparison.
\end{tcolorbox}


\subsection{Answer to RQ3: Framework Extensibility}

\begin{figure*}[th]
    \centering
    \includegraphics[width=0.8\linewidth]{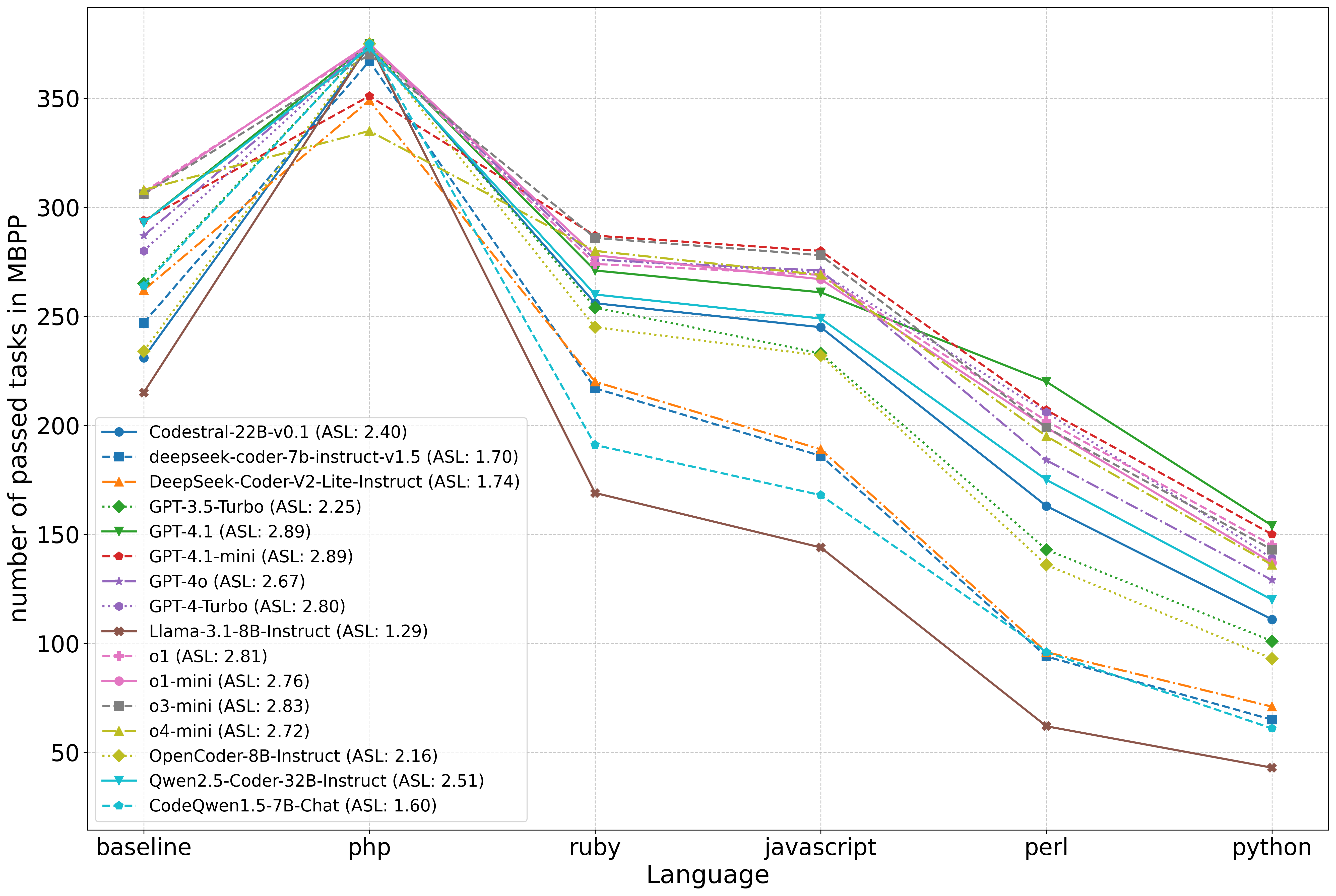}
    \caption{Robustness assessment of LLMs on Framework 2.}
    \label{lab:exp2}
\end{figure*}

To investigate the extensibility of \toolname, we construct a five-step translation chain: Python → PHP → Ruby → JavaScript → Perl → Python, creating a complete loop that returns to the original language. 
This design tests the model's ability to preserve program semantics across diverse paradigms including scripting languages, web-oriented languages, and general-purpose programming environments. 
We evaluate this extension using the MXEVAL benchmark~\cite{athiwaratkun2022multi}.
We specifically selected this benchmark due to its derivation from the original MBPP dataset, which allows us to seamlessly integrate the rigorous test suites from MBPP Plus for the Python validation phases within the loop.
By aligning the Python evaluation standard with our primary experiments, we ensure that the robustness assessment remains grounded in high-reliability correctness checks.
As shown in Fig.~\ref{lab:exp2}, the experimental results on 16 mainstream LLMs reveal substantial performance degradation across successive translation loops, with pass@1 accuracy declining by 32.00\%–54.13\% compared to the initial performance. 
Notably, the robustness rankings mirror those observed in our original framework, where initial translation performance does not necessarily correlate with sustained robustness across the complete chain. 
GPT-4.1 and GPT-4.1-mini achieve the highest ASL scores of 2.89, followed by o3-mini (2.83), demonstrating superior semantic preservation capabilities. 
Conversely, open-source models exhibit greater challenges, with Llama-3.1-8B-Instruct recording the lowest ASL score of 1.29.

\begin{tcolorbox}[
    colback=boxcolor,
    colframe=bordercolor,
    title={\textbf{Answer Summary to RQ3}},
    fonttitle=\bfseries,
    boxrule=1pt,
    arc=3pt,
    left=8pt,
    right=8pt,
    top=8pt,
    bottom=8pt
]
Our extensibility analysis reveals: (1) \textit{Broad Applicability:} \toolname successfully adapts to a complex multi-language translation chain (Python $\to$ PHP $\to \dots \to$ Python), demonstrating its flexibility beyond standard generation-summarization tasks; (2) \textit{Quantifiable Instability:} We observed substantial performance declines (32.00\%--54.13\%) across the translation loop. This confirms that \toolname effectively stress-tests model consistency across diverse software engineering paradigms, providing quantifiable robustness insights specific to cross-language adaptation scenarios.
\end{tcolorbox}

\section{Discussion}
\label{sec:discussion}

\subsection{Distinction from Prior Duality-based Evaluations}

The concept of duality in software engineering was first proposed by Wei et al~\cite{wei2019code}, which demonstrated that these complementary tasks require consistent semantic understanding from opposing perspectives. 
This foundational concept has been leveraged for different evaluation purposes.
Min et al.~\cite{min2023beyond} propose IdentityChain, a framework that evaluates whether Code LLMs can maintain semantic consistency when translating between natural language and code in both directions. 
They introduces the Test Output Match (TOM) to compares the exact outputs of two programs across all test cases rather than just pass/fail results, capturing fine-grained semantic differences by examining specific return values and complete error messages for syntax or runtime errors. 
Allamanis et al.~\cite{allamanis2024unsupervised} introduce Round-Trip Correctness (RTC) as an unsupervised evaluation method for Code LLMs that leverages the natural duality between tasks like code generation and code summarization, where a model generates a prediction, feeds it back through the inverse task, and checks if the round-trip preserves semantic equivalence to the original input. 
RTC demonstrates strong correlation with existing metrics on narrow-domain benchmarks like HumanEval~\cite{chen2021evaluating} and ARCADE~\cite{yin2022natural} while enabling evaluation across a much broader spectrum of real-world software domains without requiring costly human annotations.


\begin{table}[htbp]
\centering
\caption{Comparison of \toolname with Related Duality-based Evaluations}
\label{tab:comparison}
\begin{tabular}{@{}llll@{}}
\toprule
\textbf{Dimension} & \textbf{IdentityChain~\cite{min2023beyond}} & \textbf{RTC~\cite{allamanis2024unsupervised}} & \textbf{\toolname (Ours)} \\ \midrule
\textbf{Verification} & Output Matching (TOM) & Semantic Inference & \textbf{Test Case Execution} \\
\textbf{Mechanism} & Single Round-trip & Single Round-trip & \textbf{Iterative Stress-test} \\
\textbf{Goal} & Semantic Consistency & Unsupervised Eval & \textbf{Sustained Robustness} \\
\textbf{Incorrect Code} & \textbf{Permitted} (if consistent) & \textbf{Potential False Positive} & \textbf{Zero Tolerance} \\ \bottomrule
\end{tabular}
\end{table}

Although \toolname shares the conceptual foundation of leveraging software engineering duality (e.g., the code generation-summarization cycle) with prior works such as \textit{IdentityChain} \cite{min2023beyond} and \textit{Round-Trip Correctness} (RTC) \cite{allamanis2024unsupervised}, it represents a fundamental departure in evaluation philosophy, technical grounding, and quantitative depth. We summary their evaluation goals in Table~\ref{tab:comparison}.

\noindent\textbf{The Necessity of Correctness-anchored Stability} 
A critical limitation of existing consistency metrics is their potential detachment from functional correctness. \textit{IdentityChain} utilizes the Test Output Match (TOM) score to measure consistency. However, if a model repeatedly generates logically identical yet incorrect code (i.e., being ``consistently wrong''), it would still receive a deceptively high TOM score. Similarly, RTC operates on the unsupervised assumption that a successful round-trip (Prompt $\rightarrow$ Code $\rightarrow$ Summary) implies code correctness. Our empirical observations suggest a ``hallucination of consistency,'' where LLMs can generate accurate-looking natural language summaries for semantically flawed code. 
In contrast, \toolname establishes functional correctness as the non-negotiable bedrock of robustness. By integrating a rigorous testing phase in each loop (see details in Section~\ref{label:frame}), the evaluation immediately terminates upon a single functional failure. This design ensures that we quantify the model's ability to maintain semantic integrity through valid transformations, rather than its propensity to replicate errors.

\noindent\textbf{Static Consistency vs. Dynamic Robustness} 
While IdentityChain~\cite{min2023beyond} and RTC~\cite{allamanis2024unsupervised} work as static snapshot assessments—capturing semantic alignment within a single round-trip—\toolname introduces a dynamic stress-testing paradigm. By institutionalizing iterative duality loops, our framework enables the observation of how semantic integrity progressively decays or abruptly collapses across sustained transformations. This longitudinal perspective uncovers a critical phenomenon invisible to single-round evaluations: a model’s initial proficiency (e.g., Pass@1) is often decoupled from its sustained stability. Consequently, \toolname facilitates the identification of a "Robustness Threshold"—the precise inflection point where accumulated semantic drift triggers functional failure. This provides a granular characterization of model reliability that transcends the limitations of binary consistency metrics.

\noindent\textbf{ASL Metric vs. Binary/TOM Metrics} 
The proposed Average Sustainable Loops (ASL) metric offers a more nuanced quantification compared to traditional binary success rates or output matching ratios. 
First, the \textbf{quadratic weighting scheme} ($i^2$) recognizes and rewards long-term stability, which is essential for deploying LLMs in complex, multi-step agent systems~\cite{chen2025prometheus, bouzenia2024repairagent}. 
Second, by incorporating \textbf{LLM-based semantic similarity analysis} at the failure boundary, ASL distinguishes between failures caused by legitimate prompt evolution versus sudden model collapse under minimal perturbation. This provides developers with interpretable feedback on specific failure modes that existing metrics fail to capture.

\begin{figure}[htbp]
\centering
\begin{minipage}{0.48\textwidth}
\begin{lstlisting}[caption={Loop 1 Implementation (MBPP 119)}, label=list:loop1]
def search(nums):
    left, right = 0, len(nums) - 1
    while left < right:
        mid = left + (right - left) // 2
        if mid % 2 == 1:
            mid -= 1
        if nums[mid] == nums[mid + 1]:
            left = mid + 2
        else:
            right = mid
    return nums[left]
\end{lstlisting}
\end{minipage}
\hfill
\begin{minipage}{0.48\textwidth}
\begin{lstlisting}[caption={Loop 2 Implementation (MBPP 119)}, label=list:loop2]
def search(nums):
    low, high = 0, len(nums) - 1
    while low < high:
        mid = (low + high) // 2
        if mid % 2 == 1:
            mid -= 1
        if nums[mid] == nums[mid + 1]:
            low = mid + 2
        else:
            high = mid
    return nums[low]
\end{lstlisting}
\end{minipage}
\caption{An example of a model being ``consistently wrong'' on MBPP$^+$ No. 119 task. Both loops produce functionally equivalent code that shares the same logical flaw (failure to handle multiple unique elements), which \textit{IdentityChain} would reward as robust, but \toolname identifies as an immediate failure.}
\label{fig:consistent_wrong}
\end{figure}

\noindent\textbf{Case Study: Why Functional Anchoring Matters?} 
As illustrated in Figure~\ref{fig:consistent_wrong}, Task No. 119 from the MBPP$^+$ benchmark requires finding the unique element in a sorted array. The model generated two implementations that are semantically and functionally equivalent—both using a binary search strategy that incorrectly assumes only one unique element exists. 
Under the \textit{IdentityChain} framework, since both loops exhibit identical logic and produce matching outputs for the provided simple test cases, the model would receive a high TOM score, falsely indicating high robustness. 
In sharp contrast, \toolname executes these implementations against the rigorous MBPP Plus test suite. The evaluation detects the failure to handle complex input distributions (e.g., multiple unique elements) in the very first loop. Consequently, the loop terminates immediately, yielding an ASL score of 0. This case confirms that \toolname provides a more authentic measure of reliability by ensuring stability is anchored in correctness.

\subsection{LLM as Judge: Validation and Bias Assessment}
To quantify the robustness of our framework and ensure the integrity of the semantic similarity scores used in the ASL metric, we acknowledge that employing an LLM as an evaluator may introduce potential biases or misinterpretations. To mitigate these concerns and validate the reliability of gpt-4-turbo-2024-04-09 as our primary similarity judge, we conducted a rigorous empirical audit.

\noindent\textbf{Human Audit Protocol} We randomly sampled 100 loop transitions across various models and tasks from our evaluation corpus. Each sample—consisting of the source code, the corresponding natural language summary, and the assigned AI similarity score—was independently reviewed by three domain experts in Software Engineering. These experts were tasked with assessing two primary dimensions: 1) \textbf{Scoring Validity}: Whether the AI-generated similarity score accurately reflected the semantic equivalence (or drift) between the iterative prompts. 2) \textbf{Bias Detection}: Whether the evaluator exhibited any systematic preference for specific model architectures or failed to identify subtle functional discrepancies.

\noindent\textbf{Experimental Results} The human audit revealed a \textbf{92\% alignment rate} between the human experts and the LLM judge. In the vast majority of cases, the experts deemed the AI-generated scores to be both reasonable and unbiased, capturing the nuanced semantic shifts that occur during code-to-NL transformations. The high degree of consensus between automated scores and expert judgment confirms that \toolname's evaluation pipeline is highly stable and provides a faithful representation of model robustness. This validation supports our use of automated scoring for large-scale assessment, ensuring that the observed robustness patterns reflect genuine model capabilities rather than artifacts of the evaluation process.

\subsection{What are the Characteristics of the Failure Cases?}

\subsubsection{General Analysis}
\label{sec:general analysis}

To investigate the root causes of loop termination, we conducted an automated error attribution analysis on failure cases occurring between the 2nd and 10th loops. We define two primary failure patterns based on the interaction within the duality loop:
\begin{itemize}
    \item \textbf{Summarization Failure}: The derived prompt loses critical functional requirements, introduces ambiguity, or alters the core logic (i.e., semantic drift), rendering the task functionally different from the original intent.
    \item \textbf{Generation Failure}: The derived prompt remains functionally equivalent to the original task (even if it imposes stricter algorithmic constraints), but the model fails to generate code that passes the test cases based on this valid specification.
\end{itemize}

We employed GPT-5.2 to classify 5,487 failure cases based on a \textit{functional equivalence check}, revealing a distinct distribution where \textbf{62.7\% (3,438 cases) are attributed to Generation Failures}, while only \textbf{37.3\% (2,049 cases) stem from Summarization Failures}.
It means that for the majority of crashes, the models could effectively generate correct summarization from the code but fail to re-implement the code logic according to their self-generated summarization. This suggests that \toolname primarily stresses the model's \textit{execution consistency} rather than just semantic retention, implying that models often generate ``adversarial'' constraints for themselves---producing valid but complex prompts that exceed their own code generation capabilities, a phenomenon we further explore in Section~\ref{sec:mining_adversarial}.

\subsubsection{Mining Common Adversarial Patterns via \toolname}
\label{sec:mining_adversarial}

Beyond serving as a robust evaluation metric, \toolname functions as a generative framework for discovering \textit{common adversarial patterns}. Unlike traditional adversarial attacks that rely on synthetic perturbations (e.g., random character injection), the patterns mined within our loops are endogenously produced by the models themselves. These patterns represent realistic semantic variations that, despite their naturalness, can induce unexpected failures across a wide range of models.

\noindent\textbf{Methodology}
To systematically mine these adversarial variants, we leveraged the failure cases identified during our extensive evaluation. Specifically, we focused on a subset of 27 representative closed-source models\footnote{The selected models include Claude Opus 4, GPT-5, Claude Opus 4.1, Claude Sonnet 4.5, Claude Sonnet 4, GPT-5-Codex, o3-mini, GPT-5-mini, o3, GPT-4.1, o4-mini, GPT-4.1-mini, o1, GPT-5-nano, o1-mini, Gemini 2.5 Flash-Lite, Claude Haiku 3.5, Claude 3.7 Sonnet, Gemini 2.0 Flash, GPT-4o, Gemini 2.5 Flash, Gemini 2.5 Pro, GPT-4-Turbo, Gemini 2.0 Flash-Lite, GPT-3.5-Turbo, Claude Haiku 4.5, and Claude Haiku 3.}. First, we collected tasks that these models failed at loop 2--10. Afterward, we filtered tasks if there are at least five models failing on it. Finally, for a set of natural language prompts for each task, we retrieved the their semantic similarity to the original benchmark prompt and keep the highest one. For the filtered prompts, we excluded instances with significant semantic divergence to the original one through manual inspection. This process resulted in a final set of 46 tasks containing high-quality adversarial variants. We then used these variants to attack 30 high-performing models\footnote{The targeted models include: Claude 3.5 Haiku, Claude Opus 4, Claude Sonnet 4, DeepSeek-Coder-V2-Lite-Instruct, DeepSeek-V3, Gemini 2.0 Flash, Gemini 2.0 Flash-Lite, Gemini 2.5 Flash, Gemini 2.5 Flash-Lite, Gemini 2.5 Pro, GPT-4.1, GPT-4.1-mini, GPT-4o, GPT-4-Turbo, Llama 3.3 70B Instruct, NextCoder-14B, NextCoder-32B, o1, o3, o3-mini, o4-mini, OpenCoder-8B-Instruct, Qwen2.5-Coder-14B-Instruct, Qwen2.5-Coder-32B-Instruct, Qwen2.5-Coder-7B-Instruct, Qwen3-235B-A22B-Instruct-2507, Qwen3-30B-A3B-Instruct-2507, Qwen3-Coder-30B-A3B-Instruct, Qwen3-Coder-480B-A35B-Instruct, and Seed-Coder-8B-Instruct.} selected from our robustness ranking in RQ2.

\noindent\textbf{Results and Analysis}
To evaluate the impact of these patterns, we categorize the outcomes into four types: \textbf{Pass\_Pass} (robust), \textbf{Pass\_Fail} (successful attack), \textbf{Fail\_Pass} (improved), and \textbf{Fail\_Fail} (consistent failure). The detailed statistics for the top-20 tasks with the highest attack success rates are presented in Table~\ref{tab:adversarial_stats}. The results indicate that common semantic variations can significantly degrade model performance. For instance, in task \texttt{Mbpp/165}, 18 out of 30 top-tier models (60\%) could solve the original prompt but failed on the variant.

\begin{table}[htbp]
\centering
\caption{Performance of 30 High-Performing Models on Original vs. Mined Adversarial Prompts (Top-20 Tasks by Attack Success).}
\label{tab:adversarial_stats}
\setlength{\tabcolsep}{3pt}
\begin{tabular}{@{}lcccc|lcccc@{}}
\toprule
\textbf{Task ID} & \textbf{P\_P} & \textbf{P\_F} & \textbf{F\_P} & \textbf{F\_F} & \textbf{Task ID} & \textbf{P\_P} & \textbf{P\_F} & \textbf{F\_P} & \textbf{F\_F} \\ \midrule
Mbpp/165 & 10 & \textbf{18} & 0 & 2 & Mbpp/721 & 15 & \textbf{8} & 6 & 1 \\
Mbpp/239 & 5 & \textbf{17} & 0 & 8 & Mbpp/580 & 17 & \textbf{7} & 5 & 1 \\
Mbpp/737 & 11 & \textbf{14} & 0 & 5 & Mbpp/754 & 17 & \textbf{7} & 2 & 4 \\
Mbpp/619 & 14 & \textbf{13} & 1 & 2 & Mbpp/301 & 7 & \textbf{7} & 2 & 14 \\
Mbpp/594 & 14 & \textbf{12} & 0 & 4 & Mbpp/633 & 21 & \textbf{6} & 0 & 3 \\
Mbpp/162 & 18 & \textbf{10} & 0 & 2 & Mbpp/437 & 20 & \textbf{6} & 3 & 1 \\
Mbpp/130 & 17 & \textbf{9} & 1 & 3 & Mbpp/446 & 19 & \textbf{6} & 2 & 3 \\
Mbpp/392 & 11 & \textbf{9} & 0 & 10 & Mbpp/473 & 16 & \textbf{6} & 3 & 5 \\
Mbpp/564 & 10 & \textbf{9} & 0 & 11 & Mbpp/142 & 13 & \textbf{6} & 1 & 10 \\
Mbpp/11  & 18 & \textbf{8} & 2 & 2 & Mbpp/803 & 22 & \textbf{5} & 1 & 2 \\ \bottomrule
\multicolumn{10}{l}{\textit{Note: P\_P=Pass\_Pass, P\_F=Pass\_Fail, F\_P=Fail\_Pass, F\_F=Fail\_Fail.}}
\end{tabular}
\end{table}

\paragraph{Taxonomy of Adversarial Patterns}
To understand the mechanisms behind these failures, we categorized the 46 adversarial prompts into three distinct patterns based on the nature of the deviation from the original prompt.

\begin{itemize}
    \item \textbf{Algorithmic Constraint and Over-Specification (19 tasks):}
    This pattern occurs when the deviation prompt unnecessarily constrains the solution space by mandating specific algorithms or implementation details, rather than focusing on the functional goal. 
    For example, in \texttt{Mbpp/286}, the prompt shifts from a general request to "find the largest sum" to explicitly requiring "Kadane's algorithm." Similarly, \texttt{Mbpp/67} forces the use of a "dynamic programming table." 
    These constraints often degrade performance by forcing models into complex implementation paths where they are more prone to syntax or logic errors, or by preventing the model from utilizing built-in or optimized standard library functions.
    
    \item \textbf{Definition Shifts and Logic Traps (10 tasks):}
    These prompts introduce subtle ambiguities, remove "anchor" hints, or invert logical conditions. A prime example is \texttt{Mbpp/737}. The original prompt explicitly suggests "using regex," leading models to generate robust solutions. The adversarial variant removes this hint, causing models to revert to naive indexing approaches (e.g., checking \texttt{s[0]}), which fail on edge cases like empty strings.
    Another case is \texttt{Mbpp/580}, where the prompt changes from a negative constraint ("remove uneven elements") to a positive assertion ("extract even integers"). While logically equivalent, this reversal often trips up models that prioritize surface-level instruction following over logical deduction.
    
    \item \textbf{Semantic Rephrasing and Formalization (17 tasks):}
    Here, the core logic remains unchanged, but the prompt is rephrased using more formal, mathematical, or verbose language.
    For instance, \texttt{Mbpp/162} replaces a natural language description of a series with a formal definition involving variables and explicit summation bounds. Surprisingly, this "mathematical precision" often reduces code generation accuracy, likely because the formal notation introduces parsing ambiguity or increases the cognitive load required to map the description to code logic.
\end{itemize}

\begin{tcolorbox}[
    colback=boxcolor,
    colframe=bordercolor,
    title={\textbf{Insights derived from Failure Cases Analysis}},
    fonttitle=\bfseries,
    boxrule=1pt,
    arc=3pt,
    left=8pt,
    right=8pt,
    top=8pt,
    bottom=8pt
]
Our analysis offers counter-intuitive insights for effective prompt engineering in code generation: \textbf{(1) Avoid Implementation Micro-management:} Users should describe \textit{what} to do, not \textit{how} to do it. Forcing specific algorithms (e.g., "use Binary Search") increases the likelihood of implementation errors compared to allowing the model to select the optimal approach naturally.
\textbf{(2) Simplicity Over Formalism:} While formal specifications are precise for humans, LLMs often perform better with clear, natural language descriptions. Over-formalizing a problem (e.g., converting a list description into complex set-theoretic notation) often acts as an adversarial perturbation that degrades performance.
\end{tcolorbox}

\subsection{Implications}

\subsubsection{Defining the Scope: Criteria for Duality Task Selection}
We describe two essential criteria for further developing \toolname with more complementary tasks suitable within its framework:
\begin{itemize}
    \item \textbf{Invertibility:} The task pair must form a logical closed loop where the output of the forward task ($T_{fwd}$) serves as a valid input for the backward task ($T_{bwd}$), and vice versa. For instance, code generation (NL $\to$ Code) and summarization (Code $\to$ NL) are invertible, whereas code optimization (Code $\to$ Optimized Code) typically lacks a natural inverse that restores the non-optimized state deterministically.
    \item \textbf{Semantic Conservation:} The transformation cycle must theoretically preserve the core functional intent. In our code translation extension (RQ3), the execution logic remains invariant across languages. Task pairs involving creative expansion (e.g., ``Write a story based on this code'') violate this criterion as they introduce unbounded semantic drift, making them unsuitable for objective robustness quantification.
\end{itemize}

Practitioners can adhere to these criteria when extending \toolname to new software engineering domains to guarantee that metric fluctuations reflect genuine model instability rather than task ill-posedness.




\subsubsection{Practical Guidelines of Adapting \toolname for Different Stakeholders}
Based on our findings, we offer tailored recommendations for different actors in the AI-for-SE ecosystem:

\noindent\textbf{For Model Developers:} \toolname could be considered as a standard stress-test protocol. Unlike static benchmarks (e.g., HumanEval), our iterative loop effectively exposes ``hidden brittleness''—where models overfit to specific prompt formats but lack deep semantic grounding. Developers can utilize the failure cases (specifically the adversarial patterns mined in Section~\ref{sec:mining_adversarial}) to construct high-quality instruction-tuning datasets, thereby enhancing model alignment and structural stability.

\noindent\textbf{For Security Analysts:} The framework serves as an automated Red Teaming tool. The discovered \textit{Summarization Failures} (Section~\ref{sec:general analysis}) highlight potential attack vectors where subtle prompt rephrasing can bypass model safety guardrails or induce functional errors. Analysts can leverage the endogenously generated adversarial prompts to probe the boundaries of model reliability without manual crafting of attack samples.

\noindent\textbf{For Model Users:} When selecting foundation models for agent-based systems, users should prioritize the ASL metric over traditional Pass@1 accuracy. Our results show that high initial performance does not guarantee the long-term coherence required for autonomous agents (e.g., \textit{OpenAI o1} drops 9 positions in robustness ranking). For complex workflows where outputs are iteratively consumed, models with higher ASL scores (e.g., \textit{Qwen3} or \textit{Claude Opus}) offer significantly lower risks of cascading failure.



\subsubsection{Efficiency and Cost Analysis}
While \toolname introduces an iterative overhead compared to single-pass evaluations, the computational cost remains highly manageable and scalable.
First, the framework employs an \textit{Early Termination} mechanism: the evaluation loop halts immediately upon functional failure. Although the maximum loop depth is set to 10, our experimental results (Section 4.5) show that the average sustainable loops across all models is approximately 6.38. This indicates that the amortized computational cost is roughly $6.4\times$ that of a standard Pass@1 assessment---a modest linear factor acceptable for most research and production environments.
Second, the feasibility of large-scale deployment is empirically validated by our leaderboard construction. We successfully evaluated 96 LLMs, ranging from 0.5B to 685B parameters, using standard academic computing resources (8 NVIDIA H200 GPUs) and commercial APIs.
Given that \toolname uncovers critical reliability risks (e.g., sudden semantic collapse) that are invisible to single-pass metrics, providing unique insights to ensure the safety of software systems.

\subsection{Threats to Validity}

\textbf{External Validity.} The primary threat concerns the generalizability of our findings across different programming languages and benchmarks. 
Our large-scale evaluation of 96 models primarily utilizes the MBPP Plus benchmark~\cite{liu2024your}. 
While MBPP Plus is chosen for its rigorous quality filtering and extensive test case corpus, its tasks are predominantly self-contained Python functions. 
Consequently, the observed robustness patterns might vary in more complex, multi-file software projects or when using different programming paradigms. 
Although we demonstrated \toolname's extensibility via a code translation loop involving five languages, further studies are required to confirm if these findings hold for lower-level or domain-specific languages.

\noindent\textbf{Internal Validity.} A potential threat to internal validity arises from the reliability of our automated evaluation components. 
First, we employ GPT-4-turbo as a similarity judge to calculate the semantic scores at failure boundaries. 
To mitigate potential bias or misinterpretation, we conducted a human audit with domain experts, which revealed a 92\% alignment rate with the AI-generated scores. In the future, we intend to bolster the objectivity of our framework by implementing a cross-validation protocol that employs multiple distinct LLMs as judges, thereby mitigating potential biases inherent to single-model evaluators.
Second, our primary evaluation adopts greedy decoding to ensure experimental reproducibility. 
While our analysis in RQ3 indicates that relative model rankings remain highly stable across various temperature settings (Spearman $\rho = 0.974$), the absolute Average Sustainable Loops (ASL) scores are inherently tied to these specific decoding configurations.

\section{Conclusion}
In this paper, we addressed critical limitations in current LLM robustness evaluation for programming, which rely on biased adversarial attacks and overlook sustained consistency capabilities. We introduced \toolname, a novel framework that assesses robustness through self-contained iterative loops between dual software engineering tasks such as code generation and summarization. 
Our comprehensive evaluation of 96 LLMs reveals that initial programming proficiency does not correlate with sustained robustness, with 85 models showing significant ranking changes when evaluated for long-term stability. 
The proposed Average Sustainable Loops (ASL) metric provides a unified and unbiased robustness assessment that correlates meaningfully with software system performance while maintaining exceptional framework stability across experimental variations. \toolname offers authentic insights into model reliability for deployment scenarios requiring sustained autonomous operation, contributing to understanding of self-consistency as a fundamental dimension of LLM robustness beyond adversarial resilience.
Looking ahead, we plan to conduct qualitative analysis to investigate why certain models exhibit superior robustness while others fail quickly, and aim to develop targeted enhancement strategies that improve model stability. 

\bibliographystyle{ACM-Reference-Format}
\bibliography{sample-base}

@article{wei2019code,
  title={Code generation as a dual task of code summarization},
  author={Wei, Bolin and Li, Ge and Xia, Xin and Fu, Zhiyi and Jin, Zhi},
  journal={Advances in neural information processing systems},
  volume={32},
  year={2019}
}

@article{fang2025smaller,
  title={Smaller= Weaker? Benchmarking Robustness of Quantized LLMs in Code Generation},
  author={Fang, Sen and Ding, Weiyuan and Mastropaolo, Antonio and Xu, Bowen},
  journal={arXiv preprint arXiv:2506.22776},
  year={2025}
}

@article{shao2024deepseekmath,
  title={Deepseekmath: Pushing the limits of mathematical reasoning in open language models},
  author={Shao, Zhihong and Wang, Peiyi and Zhu, Qihao and Xu, Runxin and Song, Junxiao and Bi, Xiao and Zhang, Haowei and Zhang, Mingchuan and Li, YK and Wu, Yang and others},
  journal={arXiv preprint arXiv:2402.03300},
  year={2024}
}

@article{rafailov2023direct,
  title={Direct preference optimization: Your language model is secretly a reward model},
  author={Rafailov, Rafael and Sharma, Archit and Mitchell, Eric and Manning, Christopher D and Ermon, Stefano and Finn, Chelsea},
  journal={Advances in neural information processing systems},
  volume={36},
  pages={53728--53741},
  year={2023}
}

@article{xu2024core,
  title={Core: Llm as interpreter for natural language programming, pseudo-code programming, and flow programming of ai agents},
  author={Xu, Shuyuan and Li, Zelong and Mei, Kai and Zhang, Yongfeng},
  journal={arXiv preprint arXiv:2405.06907},
  volume={2},
  year={2024}
}

@inproceedings{oda2015learning,
  title={Learning to generate pseudo-code from source code using statistical machine translation},
  author={Oda, Yusuke and Fudaba, Hiroyuki and Neubig, Graham and Hata, Hideaki and Sakti, Sakriani and Toda, Tomoki and Nakamura, Satoshi},
  booktitle={2015 30th IEEE/ACM International Conference on Automated Software Engineering (ASE)},
  pages={574--584},
  year={2015},
  organization={IEEE}
}

@inproceedings{kwon2023efficient,
  title={Efficient memory management for large language model serving with pagedattention},
  author={Kwon, Woosuk and Li, Zhuohan and Zhuang, Siyuan and Sheng, Ying and Zheng, Lianmin and Yu, Cody Hao and Gonzalez, Joseph and Zhang, Hao and Stoica, Ion},
  booktitle={Proceedings of the 29th symposium on operating systems principles},
  pages={611--626},
  year={2023}
}

@inproceedings{
    sclar2024quantifying,
    title={Quantifying Language Models' Sensitivity to Spurious Features in Prompt Design or: How I learned to start worrying about prompt formatting},
    author={Melanie Sclar and Yejin Choi and Yulia Tsvetkov and Alane Suhr},
    booktitle={The Twelfth International Conference on Learning Representations},
    year={2024}
}

@article{roziere2023code,
  title={Code llama: Open foundation models for code},
  author={Roziere, Baptiste and Gehring, Jonas and Gloeckle, Fabian and Sootla, Sten and Gat, Itai and Tan, Xiaoqing Ellen and Adi, Yossi and Liu, Jingyu and Sauvestre, Romain and Remez, Tal and others},
  journal={arXiv preprint arXiv:2308.12950},
  year={2023}
}

@article{dubey2024llama,
  title={The llama 3 herd of models},
  author={Dubey, Abhimanyu and Jauhri, Abhinav and Pandey, Abhinav and Kadian, Abhishek and Al-Dahle, Ahmad and Letman, Aiesha and Mathur, Akhil and Schelten, Alan and Yang, Amy and Fan, Angela and others},
  journal={arXiv preprint arXiv:2407.21783},
  year={2024}
}

@article{liu2024your,
  title={Is your code generated by chatgpt really correct? rigorous evaluation of large language models for code generation},
  author={Liu, Jiawei and Xia, Chunqiu Steven and Wang, Yuyao and Zhang, Lingming},
  journal={Advances in Neural Information Processing Systems},
  volume={36},
  year={2024}
}

@article{austin2021program,
  title={Program synthesis with large language models},
  author={Austin, Jacob and Odena, Augustus and Nye, Maxwell and Bosma, Maarten and Michalewski, Henryk and Dohan, David and Jiang, Ellen and Cai, Carrie and Terry, Michael and Le, Quoc and others},
  journal={arXiv preprint arXiv:2108.07732},
  year={2021}
}

@article{chen2021evaluating,
  title={Evaluating large language models trained on code},
  author={Chen, Mark and Tworek, Jerry and Jun, Heewoo and Yuan, Qiming and Pinto, Henrique Ponde De Oliveira and Kaplan, Jared and Edwards, Harri and Burda, Yuri and Joseph, Nicholas and Brockman, Greg and others},
  journal={arXiv preprint arXiv:2107.03374},
  year={2021}
}

@inproceedings{wei2023towards,
  title={Towards greener yet powerful code generation via quantization: An empirical study},
  author={Wei, Xiaokai and Gonugondla, Sujan Kumar and Wang, Shiqi and Ahmad, Wasi and Ray, Baishakhi and Qian, Haifeng and Li, Xiaopeng and Kumar, Varun and Wang, Zijian and Tian, Yuchen and others},
  booktitle={Proceedings of the 31st ACM Joint European Software Engineering Conference and Symposium on the Foundations of Software Engineering},
  pages={224--236},
  year={2023}
}

@inproceedings{nijkamp2023codegen,
title={CodeGen: An Open Large Language Model for Code with Multi-Turn Program Synthesis},
author={Erik Nijkamp and Bo Pang and Hiroaki Hayashi and Lifu Tu and Huan Wang and Yingbo Zhou and Silvio Savarese and Caiming Xiong},
booktitle={The Eleventh International Conference on Learning Representations },
year={2023},
url={https://openreview.net/forum?id=iaYcJKpY2B_}
}

@inproceedings{jha2023codeattack,
  title={Codeattack: Code-based adversarial attacks for pre-trained programming language models},
  author={Jha, Akshita and Reddy, Chandan K},
  booktitle={Proceedings of the AAAI Conference on Artificial Intelligence},
  pages={14892--14900},
  year={2023}
}

@article{nguyen2023adversarial,
  title={Adversarial attacks on code models with discriminative graph patterns},
  author={Nguyen, Thanh-Dat and Zhou, Yang and Le, Xuan Bach D and Thongtanunam, Patanamon and Lo, David},
  journal={arXiv preprint arXiv:2308.11161},
  year={2023}
}

@article{ge2024demonstration,
  title={Demonstration Attack against In-Context Learning for Code Intelligence},
  author={Ge, Yifei and Sun, Weisong and Lou, Yihang and Fang, Chunrong and Zhang, Yiran and Li, Yiming and Zhang, Xiaofang and Liu, Yang and Zhao, Zhihong and Chen, Zhenyu},
  journal={arXiv preprint arXiv:2410.02841},
  year={2024}
}

@article{improta2025enhancing,
  title={Enhancing robustness of ai offensive code generators via data augmentation},
  author={Improta, Cristina and Liguori, Pietro and Natella, Roberto and Cukic, Bojan and Cotroneo, Domenico},
  journal={Empirical Software Engineering},
  volume={30},
  number={1},
  pages={7},
  year={2025},
  publisher={Springer}
}

@inproceedings{chen2023evaluating,
  title={Evaluating and enhancing the robustness of code pre-trained models through structure-aware adversarial samples generation},
  author={Chen, Nuo and Sun, Qiushi and Wang, Jianing and Gao, Ming and Li, Xiaoli and Li, Xiang},
  booktitle={Findings of the Association for Computational Linguistics: EMNLP 2023},
  pages={14857--14873},
  year={2023}
}

@article{yang2024robustness,
  title={Robustness, security, privacy, explainability, efficiency, and usability of large language models for code},
  author={Yang, Zhou and Sun, Zhensu and Yue, Terry Zhuo and Devanbu, Premkumar and Lo, David},
  journal={arXiv preprint arXiv:2403.07506},
  year={2024}
}

@inproceedings{zhang2024attacks,
  title={Attacks and Defenses for Large Language Models on Coding Tasks},
  author={Zhang, Chi and Wang, Zifan and Zhao, Ruoshi and Mangal, Ravi and Fredrikson, Matt and Jia, Limin and Pasareanu, Corina},
  booktitle={Proceedings of the 39th IEEE/ACM International Conference on Automated Software Engineering},
  pages={2268--2272},
  year={2024},
}

@inproceedings{mastropaolo2023robustness,
  title={On the robustness of code generation techniques: An empirical study on github copilot},
  author={Mastropaolo, Antonio and Pascarella, Luca and Guglielmi, Emanuela and Ciniselli, Matteo and Scalabrino, Simone and Oliveto, Rocco and Bavota, Gabriele},
  booktitle={2023 IEEE/ACM 45th International Conference on Software Engineering (ICSE)},
  pages={2149--2160},
  year={2023},
  organization={IEEE}
}

@article{lozhkov2024starcoder,
  title={Starcoder 2 and the stack v2: The next generation},
  author={Lozhkov, Anton and Li, Raymond and Allal, Loubna Ben and Cassano, Federico and Lamy-Poirier, Joel and Tazi, Nouamane and Tang, Ao and Pykhtar, Dmytro and Liu, Jiawei and Wei, Yuxiang and others},
  journal={arXiv preprint arXiv:2402.19173},
  year={2024}
}

@article{bavcic2024jy61,
  title={Jy61 imu sensor external validity: A framework for advanced pedometer algorithm personalisation},
  author={Ba{\v{c}}i{\'c}, Boris and Feng, Chengwei and Li, Weihua},
  journal={ISBS Proceedings Archive},
  volume={42},
  number={1},
  pages={60},
  year={2024}
}

@article{bavcic2024towards,
  title={Towards nation-wide analytical healthcare infrastructures: A privacy-preserving augmented knee rehabilitation case study},
  author={Ba{\v{c}}i{\'c}, Boris and Vasile, Claudiu and Feng, Chengwei and Ciuc{\u{a}}, Marian G},
  journal={arXiv preprint arXiv:2412.20733},
  year={2024}
}

@article{craik2002levels,
  title={Levels of processing: Past, present... and future?},
  author={Craik, Fergus IM},
  journal={Memory},
  volume={10},
  number={5-6},
  pages={305--318},
  year={2002},
  publisher={Taylor \& Francis}
}

@article{athiwaratkun2022multi,
  title={Multi-lingual evaluation of code generation models},
  author={Athiwaratkun, Ben and Gouda, Sanjay Krishna and Wang, Zijian and Li, Xiaopeng and Tian, Yuchen and Tan, Ming and Ahmad, Wasi Uddin and Wang, Shiqi and Sun, Qing and Shang, Mingyue and others},
  journal={arXiv preprint arXiv:2210.14868},
  year={2022}
}

@article{achiam2023gpt,
  title={Gpt-4 technical report},
  author={Achiam, Josh and Adler, Steven and Agarwal, Sandhini and Ahmad, Lama and Akkaya, Ilge and Aleman, Florencia Leoni and Almeida, Diogo and Altenschmidt, Janko and Altman, Sam and Anadkat, Shyamal and others},
  journal={arXiv preprint arXiv:2303.08774},
  year={2023}
}

@inproceedings{rasnayaka2024empirical,
  title={An empirical study on usage and perceptions of llms in a software engineering project},
  author={Rasnayaka, Sanka and Wang, Guanlin and Shariffdeen, Ridwan and Iyer, Ganesh Neelakanta},
  booktitle={Proceedings of the 1st International Workshop on Large Language Models for Code},
  pages={111--118},
  year={2024}
}

@inproceedings{jimenez2023swe,
  title={SWE-BENCH: CAN LANGUAGE MODELS RESOLVE REAL-WORLD GITHUB ISSUES?},
  author={Jimenez, Carlos E and Yang, John and Wettig, Alexander and Yao, Shunyu and Pei, Kexin and Press, Ofir and Narasimhan, Karthik},
  booktitle={12th International Conference on Learning Representations, ICLR 2024},
  year={2024}
}

@article{ouyang2022training,
  title={Training language models to follow instructions with human feedback},
  author={Ouyang, Long and Wu, Jeffrey and Jiang, Xu and Almeida, Diogo and Wainwright, Carroll and Mishkin, Pamela and Zhang, Chong and Agarwal, Sandhini and Slama, Katarina and Ray, Alex and others},
  journal={Advances in neural information processing systems},
  volume={35},
  pages={27730--27744},
  year={2022}
}

@article{li2022competition,
  title={Competition-level code generation with alphacode},
  author={Li, Yujia and Choi, David and Chung, Junyoung and Kushman, Nate and Schrittwieser, Julian and Leblond, R{\'e}mi and Eccles, Tom and Keeling, James and Gimeno, Felix and Dal Lago, Agustin and others},
  journal={Science},
  volume={378},
  number={6624},
  pages={1092--1097},
  year={2022},
  publisher={American Association for the Advancement of Science}
}

@inproceedings{ahmad2020transformer,
  title={A Transformer-based Approach for Source Code Summarization},
  author={Ahmad, Wasi and Chakraborty, Saikat and Ray, Baishakhi and Chang, Kai-Wei},
  booktitle={Proceedings of the 58th Annual Meeting of the Association for Computational Linguistics},
  pages={4998--5007},
  year={2020}
}

@article{liu2024deepseek,
  title={Deepseek-v3 technical report},
  author={Liu, Aixin and Feng, Bei and Xue, Bing and Wang, Bingxuan and Wu, Bochao and Lu, Chengda and Zhao, Chenggang and Deng, Chengqi and Zhang, Chenyu and Ruan, Chong and others},
  journal={arXiv preprint arXiv:2412.19437},
  year={2024}
}

@article{zaremba2025trading,
  title={Trading inference-time compute for adversarial robustness},
  author={Zaremba, Wojciech and Nitishinskaya, Evgenia and Barak, Boaz and Lin, Stephanie and Toyer, Sam and Yu, Yaodong and Dias, Rachel and Wallace, Eric and Xiao, Kai and Heidecke, Johannes and others},
  journal={arXiv preprint arXiv:2501.18841},
  year={2025}
}

@inproceedings{anand2021adversarial,
  title={Adversarial Robustness of Program Synthesis Models},
  author={Anand, Mrinal and Kayal, Pratik and Singh, Mayank},
  booktitle={Advances in Programming Languages and Neurosymbolic Systems Workshop},
  year={2021}
}

@article{min2023beyond,
  title={Beyond accuracy: Evaluating self-consistency of code large language models with identitychain},
  author={Min, Marcus J and Ding, Yangruibo and Buratti, Luca and Pujar, Saurabh and Kaiser, Gail and Jana, Suman and Ray, Baishakhi},
  journal={arXiv preprint arXiv:2310.14053},
  year={2023}
}

@article{allamanis2024unsupervised,
  title={Unsupervised evaluation of code llms with round-trip correctness},
  author={Allamanis, Miltiadis and Panthaplackel, Sheena and Yin, Pengcheng},
  journal={arXiv preprint arXiv:2402.08699},
  year={2024}
}

@article{yin2022natural,
  title={Natural language to code generation in interactive data science notebooks},
  author={Yin, Pengcheng and Li, Wen-Ding and Xiao, Kefan and Rao, Abhishek and Wen, Yeming and Shi, Kensen and Howland, Joshua and Bailey, Paige and Catasta, Michele and Michalewski, Henryk and others},
  journal={arXiv preprint arXiv:2212.09248},
  year={2022}
}

@inproceedings{zhang2024autocoderover,
  title={Autocoderover: Autonomous program improvement},
  author={Zhang, Yuntong and Ruan, Haifeng and Fan, Zhiyu and Roychoudhury, Abhik},
  booktitle={Proceedings of the 33rd ACM SIGSOFT International Symposium on Software Testing and Analysis},
  pages={1592--1604},
  year={2024}
}

@article{chen2025prometheus,
  title={Prometheus: Unified Knowledge Graphs for Issue Resolution in Multilingual Codebases},
  author={Chen, Zimin and Pan, Yue and Lu, Siyu and Xu, Jiayi and Goues, Claire Le and Monperrus, Martin and Ye, He},
  journal={arXiv preprint arXiv:2507.19942},
  year={2025}
}

@article{xia2025demystifying,
  title={Demystifying llm-based software engineering agents},
  author={Xia, Chunqiu Steven and Deng, Yinlin and Dunn, Soren and Zhang, Lingming},
  journal={Proceedings of the ACM on Software Engineering},
  volume={2},
  number={FSE},
  pages={801--824},
  year={2025},
  publisher={ACM New York, NY, USA}
}

@article{dora2025hidden,
  title={The hidden risks of LLM-generated web application code: A security-centric evaluation of code generation capabilities in large language models},
  author={Dora, Swaroop and Lunkad, Deven and Aslam, Naziya and Venkatesan, S and Shukla, Sandeep Kumar},
  journal={arXiv preprint arXiv:2504.20612},
  year={2025}
}

@article{bhatt2023purple,
  title={Purple llama cyberseceval: A secure coding benchmark for language models},
  author={Bhatt, Manish and Chennabasappa, Sahana and Nikolaidis, Cyrus and Wan, Shengye and Evtimov, Ivan and Gabi, Dominik and Song, Daniel and Ahmad, Faizan and Aschermann, Cornelius and Fontana, Lorenzo and others},
  journal={arXiv preprint arXiv:2312.04724},
  year={2023}
}

@article{bhatt2024cyberseceval,
  title={Cyberseceval 2: A wide-ranging cybersecurity evaluation suite for large language models},
  author={Bhatt, Manish and Chennabasappa, Sahana and Li, Yue and Nikolaidis, Cyrus and Song, Daniel and Wan, Shengye and Ahmad, Faizan and Aschermann, Cornelius and Chen, Yaohui and Kapil, Dhaval and others},
  journal={arXiv preprint arXiv:2404.13161},
  year={2024}
}

@article{wan2024cyberseceval,
  title={Cyberseceval 3: Advancing the evaluation of cybersecurity risks and capabilities in large language models},
  author={Wan, Shengye and Nikolaidis, Cyrus and Song, Daniel and Molnar, David and Crnkovich, James and Grace, Jayson and Bhatt, Manish and Chennabasappa, Sahana and Whitman, Spencer and Ding, Stephanie and others},
  journal={arXiv preprint arXiv:2408.01605},
  year={2024}
}

@article{mohsin2024can,
  title={Can we trust large language models generated code? a framework for in-context learning, security patterns, and code evaluations across diverse llms},
  author={Mohsin, Ahmad and Janicke, Helge and Wood, Adrian and Sarker, Iqbal H and Maglaras, Leandros and Janjua, Naeem},
  journal={arXiv preprint arXiv:2406.12513},
  year={2024}
}

@article{yan2025guiding,
  title={Guiding AI to Fix Its Own Flaws: An Empirical Study on LLM-Driven Secure Code Generation},
  author={Yan, Hao and Vaidya, Swapneel Suhas and Zhang, Xiaokuan and Yao, Ziyu},
  journal={arXiv preprint arXiv:2506.23034},
  year={2025}
}

@article{wang2022recode,
  title={ReCode: Robustness evaluation of code generation models},
  author={Wang, Shiqi and Li, Zheng and Qian, Haifeng and Yang, Chenghao and Wang, Zijian and Shang, Mingyue and Kumar, Varun and Tan, Samson and Ray, Baishakhi and Bhatia, Parminder and others},
  journal={arXiv preprint arXiv:2212.10264},
  year={2022}
}

@article{bouzenia2024repairagent,
  title={Repairagent: An autonomous, llm-based agent for program repair},
  author={Bouzenia, Islem and Devanbu, Premkumar and Pradel, Michael},
  journal={arXiv preprint arXiv:2403.17134},
  year={2024}
}

@inproceedings{honarvar2025turbulence,
  title={Turbulence: Systematically and automatically testing instruction-tuned large language models for code},
  author={Honarvar, Shahin and van der Wilk, Mark and Donaldson, Alastair F},
  booktitle={2025 IEEE Conference on Software Testing, Verification and Validation (ICST)},
  pages={80--91},
  year={2025},
  organization={IEEE}
}

@article{yefet2020adversarial,
  title={Adversarial examples for models of code},
  author={Yefet, Noam and Alon, Uri and Yahav, Eran},
  journal={Proceedings of the ACM on Programming Languages},
  volume={4},
  number={OOPSLA},
  pages={1--30},
  year={2020},
  publisher={ACM New York, NY, USA}
}

@misc{gong2026analyzing,
      title={Analyzing Message-Code Inconsistency in AI Coding Agent-Authored Pull Requests}, 
      author={Jingzhi Gong and Giovanni Pinna and Yixin Bian and Jie M. Zhang},
      year={2026},
      eprint={2601.04886},
      archivePrefix={arXiv},
      primaryClass={cs.SE},
      url={https://arxiv.org/abs/2601.04886}, 
}

@ARTICLE{10416264,
  author={Xu, Zhengkang and Guo, Shikai and Wang, Yumiao and Chen, Rong and Li, Hui and Li, Xiaochen and Jiang, He},
  journal={IEEE Transactions on Software Engineering}, 
  title={Code Comment Inconsistency Detection Based on Confidence Learning}, 
  year={2024},
  volume={50},
  number={3},
  pages={598-617},
  doi={10.1109/TSE.2024.3358489}}

@STRING{jun = "June"}










\end{document}